\documentclass[a4paper,10pt]{article}
\usepackage{ascmac}
\usepackage{mathrsfs}
\usepackage{latexsym}
\usepackage{fancybox}
\usepackage{amsmath}
\usepackage{amssymb}
\usepackage{amsthm}
\usepackage{booktabs}

\DeclareMathOperator*{\wlim}{w-lim}

\begin{document}
\title{Spectral analysis of a massless charged scalar field \\ with spacial cut-off}
\author{Kazuyuki Wada\vspace{3mm}\\ \textit{Department of Mathematics, Hokkaido University,}\\\textit{Sapporo, 060-0810, Japan}\vspace{3mm}\\ E-mail address: wadakazu@math.sci.hokudai.ac.jp}
\date{2014/5/7}
\maketitle
\begin{abstract}
The quantum system of a massless charged scalar field with a self-interaction is investigated. By introducing a spacial cut-off function, the Hamiltonian of the system is realized as a linear operator on a boson Fock space. It is proven that the Hamiltonian strongly commutes with the total charge operator. This fact implies that the state space of the charged scalar field is decomposed into the infinite direct sum of fixed total charge spaces. Moreover, under certain conditions, the Hamiltonian is bounded below, self-adjoint and has a ground ground state for an arbitrarily coupling constant. A relation between the total charge of the ground state and a number operator bound is also revealed.
\end{abstract}
\section{Introduction}
\hspace{3mm}Let us consider a quantum system of a charged scalar field $\phi(\tilde{x})$ which interacts with itself on the $1+d$ dimensional space-time $\mathbb{R}^{1+d}:=\{\tilde{x}=(x^{0}, x^{1},\dots, x^{d}): x^{\nu}\in\mathbb{R}, \nu=0,\dots, d\}$ with the Minkowski metric $g=(g_{\mu\nu})$, $g_{00}=1$, $g_{jj}=-1$, $(j=1,\dots d)$,  $g_{\mu\nu}=0$ $(\mu\neq\nu)$. The Lagrangian $\mathcal{L}$ of a complex Klein-Gordon equation with a self-interaction term is given by
\begin{equation*}
\mathcal{L}=(\partial_{\nu}\phi)(\partial^{\nu}\phi)^{\ast}-m^{2}\phi\phi^{\ast}-\frac{\lambda}{4!}(\phi\phi^{\ast})^{2},\hspace{5mm}\Big(\partial_{\nu}:=\frac{\partial}{\partial x^{\nu}},\hspace{3mm}\partial^{\nu}:=g^{\nu\rho}\partial_{\rho}\Big),
\end{equation*}
where the Einstein convention for the sum on repeated Greek indices is used, $A^{\ast}$ denotes the complex conjugate of $A$, $m\ge0$ is the mass of a particle and $\lambda>0$ is a coupling constant. Let us consider the following Lagrangian $\mathcal{L}'$:
\begin{equation}\label{lagrangian}
\mathcal{L}'=(\partial_{\nu}\phi)(\partial^{\nu}\phi)^{\ast}+\mu^{2}\phi\phi^{\ast}-\frac{\lambda}{4!}(\phi\phi^{\ast})^{2},
\end{equation} 
where $\mu>0$ is merely a parameter. $\mathcal{L}'$ is the deformation of $\mathcal{L}$ by the replacement $m^{2}\rightarrow -\mu^{2}$. As is well known, the formal quantization of $\phi$ yields particles and anti-particles. We denote by $a_{+}(k)$ (resp. $a_{-}(k)$) the formal distribution kernel of the annihilation operator for the particle (resp. anti-particle). The formal adjoint $a_{+}(k)^{\ast}$ (resp. $a_{-}(k)^{\ast}$) represents the formal distribution kernel of the creation operator for the particle (resp. anti-particle). We denote by $\phi(x)$ $(x\in\mathbb{R}^{d})$ the time-zero field of $\phi$. Then
 the Hamiltonian derived from (\ref{lagrangian}) is \textit{formally} given by,
\begin{equation}\label{formal}
H_{\textrm{formal}}=\displaystyle\int_{\mathbb{R}^{d}}|k|(a_{+}(k)^{\ast}a_{+}(k)+a_{-}(k)^{\ast}a_{-}(k))\textrm{d}k + \displaystyle\int_{\mathbb{R}^{d}}\big(-\mu^{2}\phi(x)\phi(x)^{\ast}+\frac{\lambda}{4!}(\phi(x)\phi(x)^{\ast})^{2}\big)\textrm{d}x.
\end{equation}
The integrand of second term on right hand side of (\ref{formal}) is of the form of the so-called \textit{Higgs potential}. The Lagrangian $\mathcal{L}'$ is introduced as an example of \textit{spontaneous symmetry breaking} in quantum field theory (see ,e.g., [17,19]). Unfortunately, $H_{\textrm{formal}}$ is ill-defined as a linear operator on Hilbert spaces. Therefore we need modification. 
\\
\hspace{3mm}Let $\omega$ be a non-negative function on $\mathbb{R}^{d}$ denoting a one-boson Hamiltonian. Then, the free Hamiltonian $H_{0}$ of a charged scalar field is defined by the second quantization of $\omega\oplus\omega$:
\begin{equation*}
H_{0}:=\textrm{d}\Gamma_{\textrm{b}}(\omega\oplus\omega)
\end{equation*}
on a suitable boson Fock space (see Section 2).
Let $\chi_{\textrm{sp}}$ be a non-negative function on $\mathbb{R}^{d}$ which plays a role as \textit{spacial cut-off}. For $x\in\mathbb{R}^{d}$, let $\phi(f_{x})$ be a field operator smeared by a suitable function $f_{x}$.
 The Hamiltonian $H$ under consideration is defined as follows:
 \begin{equation}\label{hamiltonian}
H:=\textrm{d}\Gamma_{\textrm{b}}(\omega\oplus\omega)+\mu\displaystyle\int_{\mathbb{R}^{d}}\chi_{\textrm{sp}}(x)\phi(f_{x})^{\ast}\phi(f_{x}) \textrm{d}x + \lambda\displaystyle\int_{\mathbb{R}^{d}}\chi_{\textrm{sp}}(x)(\phi(f_{x})^{\ast}\phi(f_{x}))^{2} \textrm{d}x,
\end{equation}
where $\mu\in\mathbb{R}$ and $\lambda>0$ are coupling constants. A rigorous definition of $H$ is introduced in Section 2. The integral on the right hand side of (\ref{hamiltonian}) is taken in the sense of strong Bochner integral. If $\mu<0$, $H$ describes a cutoff Hamiltonian of a charged scalar field with Higgs type potential. If $\mu=0$, $H$ becomes a complex-$\lambda\phi^{4}$ model with cutoffs. Hence $H$ unifies two important models. In this paper we study the properties of $H$ via operator theoretical methods. Since the interaction term of (\ref{hamiltonian}) is singular, we need careful treatment to analyze $H$. Here, \lq\lq singular" means that an interaction term is not relatively bounded with the respect to the free Hamiltonian. Introducing the spacial cut-off breaks the translation invariance of the quantum system. On the other hand, it is seen that the quantum system still holds the charge conservation. It means that the Hamiltonian $H$ and a total charge operator strongly commute. In the physical context, this property corresponds to the \textit{global U(1)-gauge symmetry}. Note that this structure is not seen in a real scalar field model. 
\\
\hspace{3mm}There are several models similar to (\ref{hamiltonian}), which have been studied so far. Glimm-Jaffe [13] considered the real $P(\phi)_{2}$ model which describes a real scalar Bose field with $\lambda\phi^{4}$-interaction in the 2-dimensional space-time. Derezi\'{n}ski-G\'{e}rard [9] considered the scattering theory for the real $P(\varphi)_{2}$ model. G\'{e}rard-Panati [12] also considered the real $P(\phi)_{2}$ model  under general settings. G\'{e}rard [11] considered the charged $P(\phi)_{2}$ model which describes the charged scalar field with a self-interaction in the 2-dimensional space-time. Note that the infimum of $\omega$ is assumed to be strictly positive in these models. An interaction model between quantum mechanical particles and a real scalar Bose field is also established. Recently, some singular perturbed models are studied. Takaesu [24] considered the generalized spin-boson model with $\phi^{4}$-perturbation. He showed the existence of a ground state and the existence of asymptotic fields for a sufficiently small coupling constant. Hidaka [15] considered the Nelson model with perturbation of a form $\sum_{j=1}^{4}c_{j}\phi^{j}$ with $c_{4}>0$. He showed the existence of a ground state for arbitrary coupling constants. A study about the total charge operator is already done by Takaesu [23] who treats a model of quantum electrodynamics. To  our best knowledge, there are few results about the charged scalar field with the infimum of $\omega$ being zero. 
\\
\hspace{3mm}We give our strategy comparing with some related works. 
\\
\hspace{6mm}\textit{Self-adjointness}: To show the self-adjointness of $H$, we apply the method in [15]. A key lemma is that the interaction term is $H$-bounded. To prove this lemma, we need the fact that the second term on the right hand side of (\ref{hamiltonian}) is infinitesimally small with respect to the third term of it. We need some technical treatments because of strong Bochner integral.
\\
\hspace{6mm}\textit{Existence of a ground state}: First of all we show the existence of a ground state of a massive Hamiltonian. After that, we consider the mass zero limit of the massive ground state. In the massive case, we apply methods used in [7,8,15] and references therein. In these methods the so-called \textit{Number-Energy Estimate} is an important lemma to show the existence of a ground state of the massive Hamiltonian. However, it is difficult to prove this lemma in our Hamiltonian since the interaction term is singular and defined by using strong Bochner integrals. As is seen below, we study the massive case without using the \textit{Number - Energy Estimate}. To show that the mass zero limit of the massive ground state is not zero, we use the methods in [17, 21] and references therein. 
\\
\hspace{6mm}\textit{Total charge of a ground state}: First, we show the strong commutativity of $H$ and the total charge operator. This fact is shown by applying properties of boson Fock space. In [23], the total charge of a ground state is studied only when coupling constants are sufficiently small. In this paper, we clarify the relation of the total charge of a ground state and a number operator bound, which contains the relevant result of [23].
\\
\hspace{3mm}This paper is organized as follows. In Section 2, we recall several notations and symbols about the abstract boson Fock space and introduce a Hamiltonian $H$ under consideration and state main results. The self-adjointness of $H$ is discussed in Section 3.  In Section 4, the spectrum of $H$ is specified. The existence of a ground state is proved in Section 5. The total charge of a ground state is discussed in Section 6. In Appendix A, some results which are used in this paper are collected. In Appendix B, we summarize the results of [2,5] which we use in Section 3 and Section 4.  
\section{A charged scalar field with spacial cut-off}
\subsection{Preliminaries}
\hspace{3mm}First of all, let us recall some notations and symbols about the abstract boson Fock space. Let $\mathscr{K}$ be a Hilbert space over $\mathbb{C}$. Then the boson Fock space over $\mathscr{K}$ is given by
\begin{equation*}
\mathscr{F}_{\textrm{b}}(\mathscr{K}):=\displaystyle\oplus_{n=0}^{\infty}\displaystyle\otimes_{\textrm{s}}^{n}\mathscr{K},
\end{equation*}
where $\displaystyle\otimes_{\textrm{s}}^{n}$ denotes the $n$-fold symmetric tensor product with $\otimes_{\textrm{s}}^{0}\mathscr{K}:=\mathbb{C}$ . The inner product is denoted by $\langle\cdot,\cdot\rangle$ which is linear in the right vector and the norm is denoted by $\|\cdot\|$. The \textit{Fock vacuum} in $\mathscr{F}_{\textrm{b}}(\mathscr{K})$ is denoted by $\Omega$ and
\begin{equation*}
\Omega:=\{1,0,0,\cdots\}\in\mathscr{F}_{\textrm{b}}(\mathscr{K}).
\end{equation*}
Let us introduce the finite particle subspace $\mathscr{F}_{\textrm{b},0}(\mathscr{K})$ as follows:
\begin{equation*}
\mathscr{F}_{\textrm{b},0}(\mathscr{K}):=\Big\{\Psi=\{\Psi^{(n)}\}_{n=0}^{\infty}\in\mathscr{F}_{\textrm{b}}(\mathscr{K}):\exists N \hspace{2mm}\text{such that}, \Psi^{(n)}=0 \hspace{2mm}\textrm{for} \hspace{2mm}\textrm{all}\hspace{2mm}n\ge N+1\Big\}.
\end{equation*}
Note that $\mathscr{F}_{\textrm{b},0}(\mathscr{K})$ is dense in $\mathscr{F}_{\textrm{b}}(\mathscr{K})$. For each $u\in \mathscr{K}$, the creation operator $A(u)^{\dag}$ is defined as follows:
\begin{equation*}
D(A(u)^{\dag}):=\Big\{\Psi=\{\Psi^{(n)}\}_{n=0}^{\infty}\in\mathscr{F}_{\textrm{b}}(\mathscr{K}):\displaystyle\sum_{n=1}^{\infty}n\big\|S_{n}(u\otimes\Psi^{(n-1)})\big\|_{\otimes_{\textrm{s}}^{n}\mathscr{K}}^{2}<\infty\Big\},
\end{equation*}
\begin{equation*}
(A(u)^{\dag}\Psi)^{(n)}:=\sqrt{n}S_{n}(u\otimes\Psi^{(n-1)}), \hspace{3mm}\Psi\in D(A(u)^{\dag}),\hspace{3mm} (n\ge1),
\end{equation*}
and $(A(u)^{\dag}\Psi)^{(0)}:=0$. Here $D(T)$ denotes the domain of a linear operator $T$ and $S_{n}$ denotes the symmetrization operator on $\otimes^{n}\mathscr{K}$. The annihilation operator with $u$ is given by the adjoint of $A(u)^{\dag}$:
\begin{equation*}
A(u):=(A(u)^{\dag})^{\ast}.
\end{equation*}
Then, for all $u,v\in\mathscr{K}$, annihilation and creation operators satisfy the following canonical commutation relations on $\mathscr{F}_{\textrm{b},0}(\mathscr{K})$:
\begin{equation*}
[A(u),A(v)]=[A(u)^{\dag},A(v)^{\dag}]=0,\hspace{5mm}[A(u),A(v)^{\dag}]=\langle u,v\rangle_{\mathscr{K}},
\end{equation*}
where $[X,Y]:=XY-YX$. For a subspace $D$ of $\mathscr{K}$, the subspace $\mathscr{F}_{\textrm{b,fin}}(D)$ is introduced as follows,
\begin{equation*}
\mathscr{F}_{\textrm{b,fin}}(D):=\textrm{L.H.}\{\Omega, A(u_{1})^{\dag}\cdots A(u_{n})^{\dag}\Omega:
n\in\mathbb{N},u_{j}\in D, j=1,\dots,n\},
\end{equation*}
where \textrm{L.H}$\{\cdots\}$ denotes the linear hull of a set $\{\cdots\}$. Note that, if $D$ is dense in $\mathscr{K}$, then $\mathscr{F}_{\textrm{b,fin}}(D)$ is dense in $\mathscr{F}_{\textrm{b}}(\mathscr{K})$.

\hspace{3mm}Let $T$ be a densely defined closable operator on $\mathscr{K}$. We denote the closure of $T$ by $\overline{T}$. Then the second quantization of $T$ is given by
\begin{equation*}
\textrm{d}\Gamma_{\textrm{b}}(T):=0\oplus\displaystyle\bigoplus_{n=1}^{\infty}\displaystyle\overline{\sum_{j=1}^{n}I\otimes\cdots\otimes I\otimes \stackrel{j-th}{T}\otimes I\cdots \otimes I\upharpoonright\hat{\otimes}^{n}_{s}D(T)},
\end{equation*} 
where $I$ is identity on $K$, $S\upharpoonright\mathcal{D}$ is the restriction of $S$ to $\mathcal{D}$ and $\hat{\otimes}_{\textrm{s}}^{n}$ denotes the $n$-fold algebraic symmetric tensor product. It is seen that $\textrm{d}\Gamma_{\textrm{b}}(T)$ is a closed operator. If $T$ is self-adjoint, so is $\textrm{d}\Gamma_{\textrm{b}}(T)$. Associated with $T$, another operator $\Gamma_{\textrm{b}}(T)$ is also defined as follows:
\begin{equation*}
\Gamma_{\textrm{b}}(T):=1\oplus\displaystyle\bigoplus_{n=1}^{\infty}\overline{T\otimes\cdots\otimes T\upharpoonright \hat{\otimes}_{\textrm{s}}^{n}D(T)}.
\end{equation*} 
Note that, if $T$ is bounded operator with operator norm $\|T\|\le1$, then $\Gamma_{\textrm{b}}(T)$ is bounded with $\|\Gamma_{\textrm{b}}(T)\|\le1$. 
\subsection{A Hamiltonian of a charged scalar field and main results}
\hspace{3mm}For a subspace $\mathcal{D}$ of a Hilbert space $\mathscr{K}$, we set 
\begin{equation*}
[\mathcal{D}]:=\mathcal{D}\oplus\mathcal{D}.
\end{equation*}
For $d\in\mathbb{N}$, the state space  $\mathscr{H}$ of a charged scalar field is given by
\begin{equation*}
\mathscr{H}:=\mathscr{F}_{\textrm{b}}([L^{2}(\mathbb{R}^{d})]),
\end{equation*}
the boson Fock space over $[L^{2}(\mathbb{R}^{d})]$. In the physical context under consideration, $[L^{2}(\mathbb{R}^{d})]$ describes the state space of a particle and an anti-particle. For $u\in L^{2}(\mathbb{R}^{d})$, the operators $a_{\pm}(u)$ and $a_{\pm}(u)^{\dag}$ on $\mathscr{H}$ are defined as follows:
\begin{equation*}
a_{+}(u):=A((u,0)),\hspace{3mm}a_{+}(u)^{\dag}:=A((u,0))^{\dag},\hspace{3mm}a_{-}(u):=A((0, u)),\hspace{3mm}a_{-}(u)^{\dag}:=A((0, u))^{\dag}.
\end{equation*}
The operators $a_{+}(u)$ and $a_{-}(u)$ are called the annihilation operator of a particle and an anti-particle with state function $u$ respectively. On the other hand, $a_{+}(u)^{\dag}$ and $a_{-}(u)^{\dag}$ are called the creation operator of a particle and an anti-particle with state function $u$ respectively. 
These operators satisfy the canonical commutation relations on the finite particle subspace  $\mathscr{F}_{\textrm{b},0}([L^{2}(\mathbb{R}^{d})])$:
\begin{equation}\label{ccr}
[a_{\pm}(u), a_{\pm}(v)]=[a_{\pm}(u),a_{\mp}(v)]=[a_{\pm}(u),a_{\mp}(v)^{\dag}]=0,\hspace{3mm}
[a_{\sharp}(u), a_{\natural}(v)^{\ast}]=\delta_{\sharp,\natural}\langle u,v\rangle_{L^{2}(\mathbb{R}^{d})},\hspace{2mm}\sharp, \natural=+\hspace{2mm}\text{or}\hspace{2mm}-.
\end{equation} 
\hspace{3mm}We denote the field operator smeared by $u\in L^{2}(\mathbb{R}^{d})$ by
\begin{equation*}
\phi(u):=\frac{1}{\sqrt{2}}(a_{+}(u)+a_{-}(u)^{\dag}).
\end{equation*} 
It is easy to see that $\phi(u)$ is densely defined and closable. We denote the closure of $\phi(u)$ by the same symbol. By von Neumann's theorem,  $\phi(u)^{\ast}\phi(u)$ and $\phi(u)\phi(u)^{\ast}$ are non-negative self-adjoint operators on $\mathscr{H}$. Note that a concrete action of $\phi(u)^{\ast}$ is as follows:
\begin{equation*}
\phi(u)^{\ast}=\frac{1}{\sqrt{2}}(a_{+}(u)^{\dag}+a_{-}(u)),\hspace{5mm}\text{on}\hspace{1mm}\mathscr{F}_{\textrm{b},0}([L^{2}(\mathbb{R}^{d})]).
\end{equation*}
By (\ref{ccr}), the field operators satisfy the following commutation relations on $\mathscr{H}_{0}$:
\begin{equation*}
[\phi(u), \phi(v)]=[\phi(u)^{\ast}, \phi(v)^{\ast}]=0,\hspace{7mm}[\phi(u), \phi(v)^{\ast}]=i\textrm{Im}\hspace{1mm}\langle u,v\rangle_{L^{2}(\mathbb{R}^{d})},
\end{equation*}
where \textrm{Im} $z$ denotes the imaginary part of $z\in\mathbb{C}$.
Let $\omega$ be the multiplication operator on $L^{2}(\mathbb{R}^{d})$ by the function
\begin{equation*}
\omega(k):=|k|,\hspace{3mm}k\in\mathbb{R}^{d}.
\end{equation*}For a linear operator $T$ on $L^{2}(\mathbb{R}^{d})$, we set $[T]:=T\oplus T$. Then the free Hamiltonian of the charged scalar field $H_{0}$ is defined by the second quantization of $[\omega]$;
\begin{equation*}
H_{0}:=\textrm{d}\Gamma_{\textrm{b}}([\omega]).
\end{equation*}
 The number operator $N_{\textrm{b}}$ is introduced as 
\begin{equation*}
N_{\textrm{b}}:=\textrm{d}\Gamma_{\textrm{b}}([1]).
\end{equation*}
For $q\in\mathbb{R}\setminus\{0\}$, the total charge operator $Q$ is defined as follows:
\begin{equation*}
Q:=\textrm{d}\Gamma_{\textrm{b}}((q\oplus-q)).
\end{equation*}
Let $\chi_{\textrm{sp}}\in L^{1}(\mathbb{R}^{d})$ be a non-negative function which plays role as a \textit{spacial cut-off}. We pick a function $\varphi$ which satisfies the following Assumption:
\vspace{2mm}
\\
\textbf{Assumption 2.1.}
$\varphi\in D(\omega^{-1/2}),\hspace{3mm}|\varphi(k)|=|\varphi(-k)|,$\hspace{3mm}\text{a.e.}$k\in\mathbb{R}^{d}$.
\vspace{2mm}
\\
\hspace{3mm}For $x\in\mathbb{R}^{d}$, $f_{x}\in L^{2}(\mathbb{R}^{d})$ is defined as follows:
\begin{equation*}
f_{x}(k):=\frac{\varphi(k)}{\sqrt{\omega(k)}}e^{-ikx},\hspace{3mm}\textrm{a.e}. \hspace{2mm}k\in\mathbb{R}^{d}.
\end{equation*}
with $kx:=k_{1}x_{1}+ \dots + k_{d}x_{d}$ for $k=(k_{1},\dots k_{d})\in\mathbb{R}^{d}$ and $x=(x_{1},\dots,x_{d})\in\mathbb{R}^{d}$. Let $\mu\in\mathbb{R}$ and $\lambda>0$ are parameters denoting coupling constants.
The Hamiltonian we study in this paper is as follows:
\begin{equation}\label{Hamiltonian}
H:=H_{0}+ \overline{\mu H_{1}+\lambda H_{2}},
\end{equation}
where
\begin{equation}\label{interact}
H_{1}:=\displaystyle\int_{\mathbb{R}^{d}}\chi_{\textrm{sp}}(x)\phi(f_{x})^{\ast}\phi(f_{x}) \textrm{d}x,\hspace{3mm}H_{2}:=\displaystyle\int_{\mathbb{R}^{d}}\chi_{\textrm{sp}}(x)\big(\phi(f_{x})^{\ast}\phi(f_{x})\big)^{2} \textrm{d}x.
\end{equation}
The integrals on the right hand sides of (\ref{interact}) are taken in the sense of $\mathscr{H}$-valued strong Bochner integral. Our first task is to find a condition for the self-adjointness of $H$. 
\vspace{2mm}
\\
\textbf{Theorem 2.1.}
\textit{Under Assumption 2.1, $H$ is bounded below, self-adjoint with $D(H)=D(H_{0})\cap D(\overline{H_{2}})$ and essentially self-adjoint on} $\mathscr{F}_{\textrm{b,fin}}([C_{0}^{\infty}(\mathbb{R}^{d})])$ \textit{for arbitrary} $\mu\in\mathbb{R}$ \textit{and} $\lambda>0$.
\vspace{2mm}
\\
\hspace{5mm}For a linear operator $T$, $\sigma(T)$ denotes the spectrum of $T$ and $\sigma_{\textrm{ess}}(T)$ denotes the \textit{essential} spectrum of $T$. If $T$ is bounded below and self-adjoint operator, then we define
\begin{equation*}
E_{0}(T):=\inf\sigma(T).
\end{equation*}
\textbf{Theorem 2.2.}
\textit{Under Assumption 2.1,}
\begin{equation*}
\sigma(H)=\sigma_{\textrm{ess}}(H)=[E_{0}(H),\infty).
\end{equation*}
\hspace{5mm}Let $T$ be a bounded below self-adjoint operator. In general, we call that $T$ has a ground state if $E_{0}(T)$ is a eigenvalue of $T$. To prove the existence of a ground state of $H$, we need the following assumption: 
\vspace{2mm}
\\
\textbf{Assumption 2.2.}
\vspace{1mm}
\\
\hspace{3mm}(1) $\varphi$ \textit{is a rotation invariant function and has a compact support}.
\vspace{2mm}
\\
\hspace{3mm}(2) \textit{There exists an open set} $\Omega\subset\mathbb{R}^{d}$ \textit{such that} $\overline{\Omega}=$\text{supp}\hspace{1mm}$\varphi$ \textit{and} $\varphi$ \textit{is continuously differentiable on} $\Omega$.
\vspace{2mm}
\\
\hspace{3mm}(3) $\varphi\in D(\omega^{-5/2})$,\hspace{3mm}$\frac{\partial\varphi}{\partial k_{j}}\in D(\omega^{-3/2})$,\hspace{3mm}$(j=1,\dots, d)$.
\vspace{2mm}
\\
\hspace{3mm}(4) $\displaystyle\int_{\mathbb{R}^{d}}(1+|x|^{2})\chi_{\textrm{sp}}(x) \textrm{d}x <\infty$.
\vspace{2mm}
\\
\textbf{Theorem 2.3.} 
\textit{Under Assumptions 2.1 and 2.2, $H$ has a ground state for arbitrary} $\mu\in\mathbb{R}$ \textit{and} $\lambda>0$.
\vspace{2mm}
\\
\hspace{5mm}The next theorem is one of characteristic structures which is not seen in the case of real scalar field and it corresponds to the charge conservation of the quantum system.  
\vspace{2mm}
\\
\textbf{Theorem 2.4.} \textit{Under Assumption 2.1}, $H$ \textit{and} $Q$ \textit{strongly commute}.
\vspace{2mm}
\\
\hspace{5mm}Let $\Phi_{g}$ be a ground state of $H$ with $\|\Phi_{g}\|=1$. By Theorem 2.4, $\mathscr{H}$ is decomposed with respect to the spectrum of the total charge $Q$ as 
\begin{equation*}
\mathscr{H}=\displaystyle\bigoplus_{z\in\mathbb{Z}}\mathscr{H}_{q}(z),
\end{equation*}  
where $\mathscr{H}_{q}(z):=$Ker$(Q-qz)$. The next result is a slight generalization of [23, Theorem1.7].
\vspace{2mm}
\\
\textbf{Theorem 2.5.} \textit{Suppose that Assumptions 2.1 and 2.2 are satisfied. Let} 
\begin{equation*}
n_{0}:=\textrm{min}\big\{n\in\mathbb{N}:\big\|N_{\textrm{b}}^{1/2}\Phi_{g}\big\|_{\mathscr{H}}^{2}<n\big\}.
\end{equation*} 
Then $\Phi_{g}\notin \mathscr{H}_{q}(z)$ for all $|z|\ge n_{0}$.
\section{Self-adjointness of $H$}
In this section, we prove Theorem 2.1. 
\vspace{2mm}
\\
\textbf{Lemma 3.1.} \textit{Assume that} $\varphi\in D(\omega^{-1/2})$, \textit{then $H$ is essentially self-adjoint on} $\mathscr{F}_{\textrm{b,fin}}([C_{0}^{\infty}(\mathbb{R}^{d}])$.
\begin{proof} First, we check that $H$ satisfies the criterion of essential self-adjointness on $D(H_{0})\cap\mathscr{F}_{\textrm{b},0}([L^{2}(\mathbb{R}^{d})])$ (see Proposition B.1). Since $\mu H_{1}+\lambda H_{2}$ maps $\otimes^{n}_{\textrm{s}}([L^{2}(\mathbb{R}^{d})])$ to $\oplus_{j=-4}^{4}\otimes^{n+j}_{\textrm{s}}([L^{2}(\mathbb{R}^{d}])$, we see that 
\begin{equation*}
\langle\Psi^{(n)},(\mu H_{1}+\lambda H_{2})\Psi^{(m)}\rangle=0,\hspace{5mm}\textrm{whenever}\hspace{3mm}|n-m|\ge 5.
\end{equation*}
 If $\mu\ge 0$, then it is obvious that $H$ is bounded below on $D(H_{0})\cap\mathscr{F}_{\textrm{b},0}([L^{2}(\mathbb{R}^{d})])$. In the case where $\mu<0$, for any $\Psi\in D(H_{0})\cap\mathscr{F}_{\textrm{b},0}([L^{2}(\mathbb{R}^{d})])$, we see that
\begin{equation*}
\begin{aligned}
\langle\Psi , H\Psi\rangle &=\langle\Psi, H_{0}\Psi\rangle+\displaystyle\int_{\mathbb{R}^{d}}\chi_{\textrm{sp}}(x)\langle \Psi, \{\mu\phi(f_{x})^{\ast}\phi(f_{x})+\lambda(\phi(f_{x})^{\ast}\phi(f_{x}))^{2}\}\Psi\rangle \textrm{d}x
\\
&\ge \displaystyle\int_{\mathbb{R}^{d}}\chi_{\textrm{sp}}(x) \textrm{d}x \displaystyle\int_{t\ge0}(\mu t+\lambda t^{2}) \textrm{d}\big\|E_{x}(t)\Psi\big\|^{2}
\\
&\ge-\frac{\mu^{2}}{4\lambda}\big\|\Psi\big\|^{2}\big\|\chi_{\textrm{sp}}\big\|_{L^{1}} >-\infty,
\end{aligned}
\end{equation*}
where $E_{x}(\cdot)$ is the spectral measure of $\phi(f_{x})^{\ast}\phi(f_{x})$. The relative boundedness of $\mu H_{1}+\lambda H_{2}$ with respect to $(N_{\textrm{b}}+1)^{2}$ is seen by using Proposition A.1. Therefore $H$ is essentially self-adjoint on $D(H_{0})\cap\mathscr{F}_{\textrm{b},0}([L^{2}(\mathbb{R}^{d})])$. Since $\mathscr{F}_{\textrm{b,fin}}([C_{0}^{\infty}(\mathbb{R}^{d})])$ is a core of $H_{0}$, for any $\Psi\in D(H_{0})\cap\mathscr{F}_{\textrm{b},0}([L^{2}(\mathbb{R}^{d})])$, there exist an $N\in \mathbb{N}$ and a sequence $\{\Psi_{n}\}_{n=1}^{\infty}$ $\subset \mathscr{F}_{\textrm{b,fin}}([C_{0}^{\infty}(\mathbb{R}^{d})])$ such that $\Psi_{n}\rightarrow \Psi$, $H_{0}\Psi_{n}\rightarrow H_{0}\Psi$ (as $n \longrightarrow \infty$) and $\Psi^{(n)}=0$ whenever $n>N$. Since $(\mu H_{1}+\lambda H_{2})\upharpoonright(\oplus_{n=0}^{N}\otimes^{n}_{s}[L^{2}(\mathbb{R}^{d})])$ is bounded , we see that $\Psi_{n}\rightarrow \Psi$ and $H\Psi_{n} \rightarrow H\Psi$. Thus the desired result follows.
\end{proof}
Let $\epsilon$ and $\eta$ be arbitrary positive constants with $\lambda^{2}-2\epsilon-\lambda^{2}\mu^{2}\eta/\epsilon>0$. Then we define a constant $C(\mu, \lambda, \epsilon, \eta)$ as follows:
\begin{equation*}
C(\mu, \lambda, \epsilon, \eta):= (\lambda^{2}-2\epsilon-\lambda^{2}\mu^{2}\eta/\epsilon)^{-1/2}\Big(\frac{\lambda^{2}\mu^{2}}{4\epsilon\eta}\big\|\chi_{\textrm{sp}}\big\|^{2}_{L^{1}}+\frac{\|\chi_{\textrm{sp}}\|_{L^{1}}^{2}}{4\epsilon}+\lambda^{2}\|\varphi\|^{4}_{L^{2}}+1\Big)^{1/2}.
\end{equation*}
\textbf{Lemma 3.2.}
\textit{Suppose that Assumption 2.1 is satisfied. Then for all} $\Psi\in D(\overline{H})$,
\begin{equation}\label{H_{1}bd}
\big\|\overline{H}_{1}\Psi\big\|\le \theta C(\mu, \lambda, \epsilon, \eta)\big\|\overline{H}\Psi\big\|+\big( \theta C(\mu, \lambda,  \epsilon, \eta)+\frac{1}{4\theta}\big)\big\|\Psi\big\|,
\end{equation}
\begin{equation}\label{H_{2}bd}
\big\|\overline{H}_{2}\Psi\big\| \le C(\mu, \lambda, \epsilon, \eta)\big(\big\|\overline{H}\Psi\big\|+\big\|\Psi\big\|\big),
\end{equation}
\textit{where $\theta$ is an arbitrary positive constant}.
\begin{proof} Since $|\varphi(k)|=|\varphi(-k)|$, we have $[\phi(f_{x}), \phi(f_{y})^{\ast}]=0$ on $\mathscr{F}_{\textrm{b},0}([L^{2}(\mathbb{R}^{d})])$ for all $x,y\in\mathbb{R}^{d}$. For any $\Psi\in \mathscr{F}_{\textrm{b,fin}}([C_{0}^{\infty}(\mathbb{R}^{d})])$, it follows that 
\begin{align}
\big\|H_{1}\Psi\big\|^{2}&=\displaystyle\iint_{\mathbb{R}^{d}\times\mathbb{R}^{d}}\chi_{\textrm{sp}}(x)\chi_{\textrm{sp}}(y)\langle\phi(f_{x})^{\ast}\phi(f_{x})\Psi, \phi(f_{y})^{\ast}\phi(f_{y})\Psi\rangle \textrm{d}x\textrm{d}y
\notag \\
&\le\displaystyle\iint_{\mathbb{R}^{d}\times\mathbb{R}^{d}}\chi_{\textrm{sp}}(x)\chi_{\textrm{sp}}(y)\big\|\Psi\big\|\big\|\phi(f_{x})^{\ast}\phi(f_{x})\phi(f_{y})^{\ast}\phi(f_{y})\Psi\big\| \textrm{d}x \textrm{d}y
\notag \\
&\le\epsilon\displaystyle\iint_{\mathbb{R}^{d}\times\mathbb{R}^{d}}\chi_{\textrm{sp}}(x)\chi_{\textrm{sp}}(y)\big\|\phi(f_{x})^{\ast}\phi(f_{x})\phi(f_{y})^{\ast}\phi(f_{y})\Psi\big\|^{2} \textrm{d}x \textrm{d}y +\frac{1}{4\epsilon}\big\|\chi_{\textrm{sp}}\big\|_{L^{1}}^{2}\big\|\Psi\big\|^{2}
\\
&=\epsilon\displaystyle\iint_{\mathbb{R}^{d}\times\mathbb{R}^{d}}\chi_{\textrm{sp}}(x)\chi_{\textrm{sp}}(y)\langle(\phi(f_{x})^{\ast}\phi(f_{x}))^{2}\Psi, (\phi(f_{y})^{\ast}\phi(f_{y}))^{2}\Psi\rangle \textrm{d}x \textrm{d}y +\frac{1}{4\epsilon}\big\|\chi_{\textrm{sp}}\big\|_{L^{1}}^{2}\big\|\Psi\big\|^{2}
\notag \\
&=\epsilon\big\|H_{2}\Psi\big\|^{2}+\frac{1}{4\epsilon}\big\|\chi_{\textrm{sp}}\big\|_{L^{1}}^{2}\big\|\Psi\big\|^{2}.
\end{align}
Here, to get (9), we used following elementary inequality:
\begin{equation}\label{element}
ab\le \epsilon a^{2}+\frac{1}{4\epsilon}b^{2},\hspace{5mm}\text{for}\hspace{2mm}a,b \ge0\hspace{2mm} \text{and} \hspace{2mm}\epsilon>0.
\end{equation}
Thus, $H_{1}$ is infinitesimally small with respect to $H_{2}$. Next we show that $H_{2}$ is $H$-bounded. For all $\Psi\in\mathscr{F}_{\textrm{b,fin}}([C_{0}^{\infty}(\mathbb{R}^{d})])$,
\begin{equation*}
\begin{aligned}
\big\|\lambda H_{2}\Psi\big\|^{2}&=\big\|(H-H_{0}-\mu H_{1})\Psi\big\|^{2}
\\
&=\big\|H\Psi\big\|^{2}-\langle H\Psi, (H_{0}+\mu H_{1})\Psi\rangle-\langle (H_{0}+\mu H_{1})\Psi, H\Psi\rangle +\big\|(H_{0}+\mu H_{1})\Psi\big\|^{2}
\\
&=\big\|H\Psi\big\|^{2}-\lambda\langle H_{2}\Psi, H_{0}\Psi\rangle-\lambda\langle H_{0}\Psi, H_{2}\Psi\rangle-2\lambda\mu\textrm{Re}\langle H_{1}\Psi, H_{2}\Psi\rangle - \big\|(H_{0}+\mu H_{1})\Psi\big\|^{2}
\\
&\le\big\|H\Psi\big\|^{2}-\lambda\langle H_{2}\Psi, H_{0}\Psi\rangle-\lambda\langle H_{0}\Psi, H_{2}\Psi\rangle+2\lambda|\mu||\textrm{Re}\langle H_{1}\Psi, H_{2}\Psi\rangle|,
\end{aligned}
\end{equation*}
where Re $z$ denotes the real part of $z\in\mathbb{C}$. By using (10) and (\ref{element})$, 2\lambda|\mu||\textrm{Re}\langle H_{1}\Psi,H_{2}\Psi\rangle|$ is estimated as follows:
\begin{equation*}
\begin{aligned}
2\lambda|\mu||\textrm{Re}\langle H_{1}\Psi,H_{2}\Psi\rangle|&\le2\lambda|\mu|\big\|H_{1}\Psi\big\|\big\|H_{2}\Psi\big\|
\\
&\le \epsilon\big\|H_{2}\Psi\big\|^{2}+\frac{\lambda^{2}\mu^{2}}{\epsilon}\big\|H_{1}\Psi\big\|^{2}
\\
&\le \epsilon\big\|H_{2}\Psi\big\|^{2}+\frac{\lambda^{2}\mu^{2}}{\epsilon}\Big(\eta\big\|H_{2}\Psi\big\|^{2}+\frac{1}{4\eta}\big\|\chi_{\textrm{sp}}\big\|^{2}_{L^{1}}\big\|\Psi\big\|^{2}\Big),
\end{aligned}
\end{equation*}
where $\epsilon$ and $\eta$ are arbitrary positive constants. Therefore we have 
\begin{equation*}
\big\|\lambda H_{2}\Psi\big\|^{2}\le \big\|H\Psi\big\|^{2}-\lambda\displaystyle\int_{\mathbb{R}^{d}}\chi_{\textrm{sp}}(x)\langle \Psi, \{(\phi(f_{x})^{\ast})\phi(f_{x}))^{2}, H_{0}\}\Psi\rangle \textrm{d}x +(\epsilon+\frac{\lambda^{2}\mu^{2}\eta}{\epsilon})\big\|H_{2}\Psi\big\|^{2}+\frac{\lambda^{2}\mu^{2}}{4\epsilon\eta}\big\|\chi_{\textrm{sp}}\big\|^{2}_{L^{1}}\big\|\Psi\big\|^{2},
\end{equation*} 
where $\{X, Y\}:=XY+YX$. By using the identity $X^{2}Y+YX^{2}=2XYX+[X,[X,Y]]$ and the positivity of $H_{0}$, we see that
\begin{equation*}
\big\|\lambda H_{2}\Psi\big\|^{2}\le \big\|H\Psi\big\|^{2}-\lambda\displaystyle\int_{\mathbb{R}^{d}}\chi_{\textrm{sp}}(x)\langle \Psi, [\phi(f_{x})^{\ast}\phi(f_{x}), [\phi(f_{x})^{\ast}\phi(f_{x}), H_{0}]]\Psi\rangle \textrm{d}x +(\epsilon+\frac{\lambda^{2}\mu^{2}\eta}{\epsilon})\big\|H_{2}\Psi\big\|^{2}+\frac{\lambda^{2}\mu^{2}}{4\epsilon\eta}\big\|\chi_{\textrm{sp}}\big\|^{2}_{L^{1}}\big\|\Psi\big\|^{2},
\end{equation*}
 By applying Proposition A.2, we have
\begin{equation*}
[\phi(f_{x})^{\ast}\phi(f_{x}),[\phi(f_{x})^{\ast}\phi(f_{x}), \textrm{d}\Gamma_{\textrm{b}}([\omega])]]=-2\big\|\varphi\big\|_{L^{2}}^{2}\phi(f_{x})^{\ast}\phi(f_{x}).
\end{equation*}
Hence it follows that
\begin{equation}\label{abc}
\big\|\lambda H_{2}\Psi\big\|^{2}\le\big\|H\Psi\big\|^{2}+2\lambda\big\|\varphi\big\|^{2}_{L^{2}}\langle\Psi, H_{1}\Psi\rangle+(\epsilon+\frac{\lambda^{2}\mu^{2}\eta}{\epsilon})\big\|H_{2}\Psi\big\|^{2}+\frac{\lambda^{2}\mu^{2}}{4\epsilon\eta}\big\|\chi_{\textrm{sp}}\big\|^{2}_{L^{1}}\big\|\Psi\big\|^{2} .
\end{equation}
By using (10) and (\ref{element}), we have
\begin{align}
2\lambda\big\|\varphi\big\|_{L^{2}}^{2}\langle \Psi, H_{1}\Psi\rangle&\le 2\lambda\|\varphi\|_{L^{2}}^{2}\|\Psi\|\|H_{1}\Psi\|
\notag \\
&\le\big\|H_{1}\Psi\big\|^{2}+\lambda^{2}\|\varphi\|_{L^{2}}^{4}\|\Psi\|^{2}
\notag \\
&\le \epsilon\big\|H_{2}\Psi\big\|^{2}+\Big(\frac{\|\chi_{\textrm{sp}}\|_{L^{1}}^{2}}{4\epsilon}+\lambda^{2}\|\varphi\|^{4}_{L^{2}}\Big)\|\Psi\|^{2}.
\end{align}
From (12) and (13), it is seen that
\begin{equation*}
\big\|\lambda H_{2}\Psi\big\|^{2}\le \big\|H\Psi\big\|^{2}+\Big(2\epsilon+\frac{\lambda^{2}\mu^{2}\eta}{\epsilon}\Big)\big\|H_{2}\Psi\big\|+\Big(\frac{\lambda^{2}\mu^{2}}{4\epsilon\eta}\big\|\chi_{\textrm{sp}}\big\|^{2}_{L^{1}}+\frac{\|\chi_{\textrm{sp}}\|_{L^{1}}^{2}}{4\epsilon}+\lambda^{2}\|\varphi\|^{4}_{L^{2}}\Big)\big\|\Psi\big\|^{2}.
\end{equation*}
Thus, by choosing constants  $\epsilon$ and $\eta$ such that $2\epsilon+\lambda^{2}\mu^{2}\eta/\epsilon<\lambda^{2}$, we have following inequality:
\begin{equation*}
(\lambda^{2}-2\epsilon-\lambda^{2}\mu^{2}\eta/\epsilon)\big\|H_{2}\Psi\big\|^{2}\le\big\|H\Psi\big\|^{2}+\Big(\frac{\lambda^{2}\mu^{2}}{4\epsilon\eta}\big\|\chi_{\textrm{sp}}\big\|^{2}_{L^{1}}+\frac{\|\chi_{\textrm{sp}}\|_{L^{1}}^{2}}{4\epsilon}+\lambda^{2}\|\varphi\|^{4}_{L^{2}} \Big)\big\|\Psi\big\|^{2}.
\end{equation*}
Thus (\ref{H_{2}bd}) holds for all $\Psi\in\mathscr{F}_{\textrm{b,fin}}([C_{0}^{\infty}(\mathbb{R}^{d})])$.
Since $\mathscr{F}_{\textrm{b,fin}}([C_{0}^{\infty}(\mathbb{R}^{d})])$ is a core of $\overline{H}$,  (\ref{H_{2}bd}) follows for all $\Psi\in D(\overline{H})$ from a limiting argument. (\ref{H_{1}bd}) immediately follows from (\ref{H_{2}bd}) and (\ref{element}).
\end{proof}
\begin{proof}[Proof of Theorem 2.1]
\hspace{3mm}It suffices to show that $D(\overline{H})\subset D(H)$. For any $\Psi\in D(\overline{H})$, there exists a sequence $\{\Psi_{n}\}_{n=1}^{\infty}\subset \mathscr{F}_{\textrm{b,fin}}([C_{0}^{\infty}(\mathbb{R}^{d})])$ such that
\begin{equation*}
\Psi_{n}\rightarrow \Psi,\hspace{5mm}H\Psi_{n}\rightarrow \overline{H}\Psi,\hspace{5mm}(\textrm{as}\hspace{2mm}n\rightarrow\infty).
\end{equation*}
By Lemma 3.2, $H_{0}$ is $H$-bounded on $\mathscr{F}_{\textrm{b,fin}}([C_{0}^{\infty}(\mathbb{R}^{d})])$. Indeed we note that following inequality holds:
\begin{equation*}
\big\|H_{0}\Psi\big\|=\big\|(H-\mu H_{1}-\lambda H_{2})\Psi\big\|\le\big\|H\Psi\big\|+|\mu|\big\|H_{1}\Psi\big\|+\lambda\big\|H_{2}\Psi\big\|.
\end{equation*}
Therefore, $\{H_{0}\Psi_{n}\}_{n=1}^{\infty}$ and $\{H_{2}\Psi_{n}\}_{n=1}^{\infty}$ are Cauchy sequences. By the closedness of $H_{0}$ and the closability of $H_{2}$, it follows that $\Psi\in D(H_{0})\cap D(\overline{H}_{2})=D(H)$.
\end{proof}
\section{Identification of $\sigma(H)$}
In this section, we prove Theorem 2.2. Throughout this section, we always assume Assumption 2.1. Let us calculate $[\mu H_{1}+\lambda H_{2}, A((u, v))^{\dag}]$ with $u,v\in\mathscr{H}$. For all $\Psi\in\mathscr{F}_{\textrm{b,fin}}([C_{0}^{\infty}(\mathbb{R}^{d})])$, we see that
\begin{equation*}
\big[\mu H_{1} + \lambda H_{2} , A((u, v))^{\dag}\big]\Psi =\frac{1}{\sqrt{2}}(\mu T_{1}+\mu T_{2}+ 2\lambda T_{3}+ 2\lambda T_{4})\Psi,
\end{equation*}
where,
\\ 
\hspace{15mm}
$T_{1}:=\displaystyle\int_{\mathbb{R}^{d}}\chi_{\textrm{sp}}(x)\langle f_{x}, v\rangle \phi(f_{x}) \textrm{d}x,$  \hspace{24mm}
$T_{2}:=\displaystyle\int_{\mathbb{R}^{d}}\chi_{\textrm{sp}}(x)\langle f_{x}, u\rangle \phi(f_{x})^{\ast} \textrm{d}x, $ 
\\
\hspace{15mm}
$T_{3}:=\displaystyle\int_{\mathbb{R}^{d}}\chi_{\textrm{sp}}(x)\langle f_{x}, v\rangle\phi(f_{x})\phi(f_{x})^{\ast}\phi(f_{x}) \textrm{d}x, $
\hspace{5mm}
$T_{4}:=\displaystyle\int_{\mathbb{R}^{d}}\chi_{sp}(x)\langle f_{x}, u\rangle\phi(f_{x})^{\ast}\phi(f_{x})\phi(f_{x})^{\ast}\textrm{d}x $.
\vspace{3mm}
\\
Note that integrals of right hand side are taken in the $\mathscr{H}$-valued strong Bochner integral. 
\vspace{2mm}
\\
\textbf{Lemma 4.1.} $T_{j}$ ($j=1,2,3,4$) \textit{are} $H$\textit{-bounded on} $\mathscr{F}_{\text{b,fin}}([C_{0}^{\infty}(\mathbb{R}^{d})])$.
\begin{proof}Let $\Psi\in\mathscr{F}_{\textrm{b,fin}}([C_{0}^{\infty}(\mathbb{R}^{d})])$. Then 
\begin{equation*}
 \begin{aligned}
\big\|T_{1}\Psi\big\|^{2}&\le\displaystyle\int_{\mathbb{R}^{d}\times\mathbb{R}^{d}}\chi_{\textrm{sp}}(x)\chi_{\textrm{sp}}(y)|\langle f_{x}, v\rangle\langle f_{y},v\rangle|\langle\Psi, \phi(f_{y})^{\ast}\phi(f_{x})\Psi\rangle| \textrm{d}x \textrm{d}y 
\\
&\le \frac{1}{2}\big\|\omega^{-1/2}\varphi\big\|^{2}\big\|v\big\|^{2}\displaystyle\int_{\mathbb{R}^{d}\times\mathbb{R}^{d}}\chi_{\textrm{sp}}(x)\chi_{\textrm{sp}}(y)\langle\phi(f_{y})^{\ast}\phi(f_{y})\Psi\, \phi(f_{x})^{\ast}\phi(f_{x})\Psi\rangle \textrm{d}x \textrm{d}y +\frac{1}{2}\big\|\omega^{-1/2}\varphi\big\|^{2}\big\|v\big\|^{2}\big\|\chi_{\textrm{sp}}\big\|^{2}_{L^{1}}\big\|\Psi\big\|^{2}
\\
&=\frac{1}{2}\big\|\omega^{-1/2}\varphi\big\|^{2}\big\|v\big\|^{2}\big(\big\|H_{1}\Psi\big\|^{2}+\big\|\chi_{\textrm{sp}}\big\|_{L^{1}}^{2}\big\|\Psi\big\|^{2}\big).
\end{aligned}
\end{equation*} 
By applying Lemma 3.2, $T_{1}$ is $H$-bounded. It is shown that $T_{2}$ is also $H$-bounded. Next, we show the $H$-boundedness of $T_{3}$. It follows that
\begin{equation*}
 \begin{aligned}
\big\|T_{3}\Psi\big\|^{2}&\le\displaystyle\int_{\mathbb{R}^{d}\times\mathbb{R}^{d}}\chi_{\textrm{sp}}(x)\chi_{\textrm{sp}}(y)\big|\langle f_{x}, v\rangle\big|\big|\langle f_{y}, v\rangle\big|\big|\langle  \phi(f_{y})^{\ast}\phi(f_{x})\Psi,\phi(f_{x})^{\ast}\phi(f_{x})\phi(f_{y})^{\ast}\phi(f_{y})\Psi \rangle\big|\textrm{d}x\textrm{d}y
\\
&\le\frac{1}{2}\big\|\omega^{-1/2}\varphi\big\|^{2}\big\|v\big\|^{2}\displaystyle\int_{\mathbb{R}^{d}\times\mathbb{R}^{d}}\chi_{\textrm{sp}}(x)\chi_{\textrm{sp}}(y)\langle \phi(f_{y})^{\ast}\phi(f_{x})\Psi, \phi(f_{y})^{\ast}\phi(f_{x})\Psi\rangle \textrm{d}x\textrm{d}y
 \\
&\hspace{10mm}+\frac{1}{2}\big\|\omega^{-1/2}\varphi\big\|^{2}\big\|v\big\|^{2}\displaystyle\int_{\mathbb{R}^{d}\times\mathbb{R}^{d}}\chi_{\textrm{sp}}(x)\chi_{\textrm{sp}}(y) \langle\phi(f_{x})^{\ast}\phi(f_{x})\phi(f_{y})^{\ast}\phi(f_{y})\Psi, \phi(f_{x})^{\ast}\phi(f_{x})\phi(f_{y})^{\ast}\phi(f_{y})\Psi \rangle \textrm{d}x\textrm{d}y
\\
&=\frac{1}{2}\big\|\omega^{-1/2}\varphi\big\|^{2}\big\|v\big\|^{2}\big(\big\|H_{1}\Psi\big\|^{2}+\big\|H_{2}\Psi\big\|^{2}\big).
 \end{aligned}
\end{equation*} 
Thus $T_{3}$ is $H$-bounded by Lemma 3.2. The case of $T_{4}$ is also estimated similarly. Thus the desired results follow.
\end{proof}
Let $\{u_{n}\}_{n=1}^{\infty}$ and $\{v_{n}\}_{n=1}^{\infty} \subset D(\omega)\cap D(\omega^{-1/2})$ be arbitrary sequences such that
\begin{equation*}
 \wlim_{n\rightarrow\infty}u_{n}=0,\hspace{3mm}\wlim_{n\rightarrow\infty}v_{n}=0\hspace{3mm} \text{and} \hspace{2mm}\|u_{n}\|^{2}+\|v_{n}\|^{2}=1,\hspace{2mm}(n\in \mathbb{N}),
\end{equation*}
where $\wlim$ denotes a weak limit. 
It is seen that 
\begin{equation*}
\begin{aligned}
\mathscr{F}_{\textrm{b,fin}}([C_{0}^{\infty}(\mathbb{R}^{d})])&\subset D((\mu H_{1}+\lambda H_{2})A((u_{n}, v_{n}))^{\dag})\cap D(A((u_{n}, v_{n}))^{\dag}(\mu H_{1}+\lambda H_{2}))
\\
&\hspace{30mm}\cap D((\mu H_{1}+\lambda H_{2})^{\ast}A((u_{n}, v_{n})))\cap D(A((u_{n}, v_{n}))(\mu H_{1}+\lambda H_{2})^{\ast}).
\end{aligned}
\end{equation*}
By applying Proposition B.3 as $A=\mu H_{1}+\lambda H_{2}$, $B= A((u_{n}, v_{n}))^{\dag}$, $C=H$ and  $\mathcal{D}=\mathcal{E}_{C}=\mathcal{F}_{\textrm{b,fin}}([C_{0}^{\infty}(\mathbb{R}^{d})])$, we see that the weak commutator $[\mu H_{1}+\lambda H_{2}, A((u_{n}, v_{n}))]_{\textrm{w},D(H)}$ exists and
\begin{equation}\label{spec}
[\mu H_{1}+\lambda H_{2}, A((u_{n}, v_{n}))^{\dag}]_{\textrm{w},D(H)}=\frac{1}{\sqrt{2}}\Big(\overline{\mu T_{1,n}}+\overline{\mu T_{2,n}}+\overline{2\lambda T_{3,n}}+\overline{2\lambda T_{4,n}}\Big)\upharpoonright D(H),
\end{equation}
where
\\ 
\hspace{15mm}
$T_{1.n}:=\displaystyle\int_{\mathbb{R}^{d}}\chi_{\textrm{sp}}(x)\langle f_{x}, v_{n}\rangle \phi(f_{x}) \textrm{d}x$  \hspace{27mm}
$T_{2.n}:=\displaystyle\int_{\mathbb{R}^{d}}\chi_{\textrm{sp}}(x)\langle f_{x}, u_{n}\rangle \phi(f_{x})^{\ast} \textrm{d}x $ 
\\
\hspace{15mm}
$T_{3.n}:=\displaystyle\int_{\mathbb{R}^{d}}\chi_{\textrm{sp}}(x)\langle f_{x}, v_{n}\rangle\phi(f_{x})\phi(f_{x})^{\ast}\phi(f_{x}) \textrm{d}x $
\hspace{5mm}
$T_{4.n}:=\displaystyle\int_{\mathbb{R}^{d}}\chi_{sp}(x)\langle f_{x}, u_{n}\rangle\phi(f_{x})^{\ast}\phi(f_{x})\phi(f_{x})^{\ast} \textrm{d}x $.
\begin{proof}[Proof of Theorem 2.2]
We apply Proposition B.4. Hence we need only to show that for all $\Psi\in D(H)$,
\begin{equation*}
\displaystyle\lim_{n\rightarrow\infty} [\mu H_{1}+\lambda H_{2}, A((u_{n}, v_{n}))^{\dag}]_{\textrm{w}, D(H)}\Psi =0.
\end{equation*}
By (\ref{spec}), we have 
\begin{equation*}
\displaystyle\lim_{n\rightarrow\infty}[\mu H_{1}+\lambda H_{2}, A((u_{n}, v_{n}))]_{\textrm{w},D(H)}\Psi= \displaystyle\lim_{n\rightarrow\infty}\frac{1}{\sqrt{2}}(\mu \overline{T_{1,n}}+\mu \overline{T_{2,n}}+2\lambda \overline{T_{3,n}}+2\lambda \overline{T_{4,n}})\Psi.
\end{equation*}
Thus it suffices to show that $\lim_{n\rightarrow\infty}\|\overline{T_{j,n}}\Psi\|=0\hspace{2mm}(j=1,2,3,4)$. First, we consider $T_{1,n}$. Since $\mathscr{F}_{\textrm{b,fin}}([C_{0}^{\infty}(\mathbb{R}^{d})])$ is a core of $H$, there exists a sequence $\{\Psi_{k}\}_{k}\subset\mathscr{F}_{\textrm{b,fin}}([C_{0}^{\infty}(\mathbb{R}^{d})])$ such that $\Psi_{k}\rightarrow\Psi$, $H\Psi_{k}\rightarrow H\Psi\hspace{2mm}(k\rightarrow \infty)$. Then $T_{1,n}\Psi_{k}\rightarrow \overline{T_{1,n}}\Psi\hspace{2mm}(k\rightarrow\infty)$ and, for any $k\in\mathbb{N}$, we have
\begin{equation*}
\begin{aligned}
\big\|\overline{T_{1,n}}\Psi\big\|&\le\big\|\overline{T_{1,n}}\Psi-T_{1,n}\Psi_{k}\big\|+\big\|T_{1,n}\Psi_{k}\big\|
\\
&\le C\big\|H(\Psi-\Psi_{k})\big\|+D\big\|\Psi-\Psi_{k}\big\|+E\big\|(N_{\textrm{b}}+1)^{1/2}\Psi_{k}\big\|\displaystyle\int_{\mathbb{R}^{d}}\chi_{\textrm{sp}}(x)|\langle f_{x}, v_{n}\rangle| \textrm{d}x,
\end{aligned}
\end{equation*}
where $C$, $D$ and $E$ are positive constants independent of $n$ and $k$. By the property of $v_{n}$, it follows that 
\begin{equation*}
\displaystyle\lim_{n\rightarrow\infty}|\langle f_{x}, v_{n}\rangle|=0,\hspace{3mm}\text{for }x\in\mathbb{R}^{d},
\end{equation*}
and
\begin{equation*}
\chi_{\textrm{sp}}(x)|\langle f_{x}, v_{n}\rangle|\le \chi_{\textrm{sp}}(x) \big\|\omega^{-1/2}\varphi\big\|_{L^{2}}
\end{equation*}
is integrable. 
Hence, by applying the Lebesgue dominated convergence theorem, we have
\begin{equation*}
\limsup_{n\rightarrow\infty}\big\|\overline{T_{1,n}}\Psi\big\|\le  C\big\|\overline{H}(\Psi-\Psi_{k})\big\|+D\big\|\Psi-\Psi_{k}\big\|.
\end{equation*}
Since $k\in\mathbb{N}$ is arbitrary, we have $\lim_{n\rightarrow\infty}\|\overline{T_{1,n}}\Psi\|=0$ by taking $k\rightarrow\infty$. In the same manner, we can show that $\lim_{n\rightarrow\infty}\|\overline{T_{j,n}}\Psi\|=0\hspace{2mm}(j=2,3,4)$.
\end{proof}
\section{Existence of a ground state}
In this section, we prove Theorem 2.3. Throughout this section, we always suppose that Assumption 2.1 holds. For a positive constant $m>0$, we define $\omega_{m}(k)$ by
\begin{equation*}
\omega_{m}(k):=\sqrt{k^{2}+m^{2}},\hspace{5mm}k\in\mathbb{R}^{d}.
\end{equation*}
The constant $m>0$ is regarded as the mass of a boson. Let us introduce a \textit{massive} Hamiltonian $H_{m}$ as follows:
\begin{equation*}
H_{m}:=\textrm{d}\Gamma_{\textrm{b}}([\omega_{m}])+\overline{\mu H_{1}+\lambda H_{2}}.
\end{equation*}
In the same way as in the proof of Theorem 2.1, one can show that $H_{m}$ is self-adjoint, bounded below and essentially self-adjoint on $\mathscr{F}_{\textrm{b,fin}}([C_{0}^{\infty}(\mathbb{R}^{d})])$.
\vspace{2mm}
\\
\textbf{Remark 5.1.} 
The operators $H_{1}$ and $H_{2}$ are $H_{m}$-bounded with
\begin{equation*}
\big\|\overline{H}_{1}\Psi\big\|\le \theta C_{m}(\mu, \lambda, \epsilon, \eta)\big\|H_{m}\Psi\big\|+\big( \theta C_{m}(\mu, \lambda,  \epsilon, \eta)+\frac{1}{4\theta}\big)\big\|\Psi\big\|,
\end{equation*}
\begin{equation*}
\big\|\overline{H}_{2}\Psi\big\| \le C_{m}(\mu, \lambda, \epsilon, \eta)\big(\big\|H_{m}\Psi\big\|+\big\|\Psi\big\|\big),\hspace{5mm}\Psi\in D(H_{m}),
\end{equation*}
where $\theta$ is arbitrary positive constant and 
\begin{equation*}
C_{m}(\mu, \lambda, \epsilon, \eta):=(\lambda^{2}-2\epsilon-\lambda^{2}\mu^{2}\eta/\epsilon)^{-1/2}\Big(\frac{\lambda^{2}\mu^{2}}{4\epsilon\eta}\big\|\chi_{\textrm{sp}}\big\|^{2}_{L^{1}}+\frac{\|\chi_{\textrm{sp}}\|_{L^{1}}^{2}}{4\epsilon}+\lambda^{2}\|\omega_{m}^{1/2}\omega^{-1/2}\varphi\|^{4}_{L^{2}}+1\Big)^{1/2},
\end{equation*}
with $\epsilon>0$ and $\eta>0$ being arbitrary such that $\lambda^{2}>2\epsilon+\lambda^{2}\mu^{2}\eta/\epsilon$. Therefore $\textrm{d}\Gamma_{\textrm{b}}([\omega_{m}])$ is $H_{m}$-bounded.
\vspace{2mm}
\\
Let us consider the extended Hilbert space $\mathscr{H}^{\textrm{e}}$ defined by
\begin{equation*}
\mathscr{H}^{\textrm{e}}:=\mathscr{H}\otimes\mathscr{H}.
\end{equation*}
Then the \textit{extended Hamiltonian} $H_{m}^{\textrm{e}}$ is defined as follows:
\begin{equation*}
H^{\textrm{e}}_{m}:= H_{m}\otimes1_{\mathscr{H}} + 1_{\mathscr{H}}\otimes \textrm{d}\Gamma_{\textrm{b}}([\omega_{m}]),
\end{equation*}
\begin{equation*}
H_{0,m}^{\textrm{e}}:=\textrm{d}\Gamma_{\textrm{b}}([\omega_{m}])\otimes1_{\mathscr{H}}+1_{\mathscr{H}}\otimes \textrm{d}\Gamma_{\textrm{b}}([\omega_{m}]).
\end{equation*}
Let us introduce a partition of unity. Let $j_{0}, j_{\infty}$ be $\mathbb{R}$-valued functions such that $j_{0}$, $j_{\infty}\in C^{\infty}(\mathbb{R}^{d})$, $j_{0}^{2}+j_{\infty}^{2}\equiv 1$, $0\le j_{0}, j_{\infty}\le1$ and
\begin{equation*}
j_{0}(x)=   \begin{cases}
               1 & \text{$|x|\le1 $}, \\
               0 & \text{$|x| \ge 2$}.
             \end{cases}
\end{equation*}
\vspace{3mm}
\\
We set for $R>0$, $j_{0,R}:=j_{0}(\cdot/R), j_{\infty,R}:=j_{\infty}(\cdot/R)$ and $\hat{j}_{0,R}:=j_{0,R}(-i\nabla_{k})$, $\hat{j}_{\infty,R}:=j_{\infty,R}(-i\nabla_{k})$, where $\nabla_{k}:=(\partial/\partial k_{1}, \dots, \partial/\partial k_{d})$. We introduce an operator $\hat{j}_{R}$ which acts on $\oplus^{2}L^{2}(\mathbb{R}^{d})$ to $\oplus^{4}L^{2}(\mathbb{R}^{d})$ as follows:
 \begin{equation*}
\hat{j}_{R}(u, v):=\big(\hat{j}_{0,R}u,\hspace{1mm}\hat{j}_{0,R}v,\hspace{1mm}\hat{j}_{\infty,R}u,\hspace{1mm} \hat{j}_{\infty,R}v\big),
\hspace{5mm}(u, v)\in [L^{2}(\mathbb{R}^{d})].
\end{equation*}
Note that $\hat{j}_{R}$ is isometry. Let us denote the unitary operator which maps  $\mathscr{F}_{\textrm{b}}(\oplus^{4}L^{2}(\mathbb{R}^{d}))$ to $\mathscr{H}^{\textrm{e}}$ by  $U_{[L^{2}(\mathbb{R}^{d})], [L^{2}(\mathbb{R}^{d})]}$ (see Proposition A.3). 
We define an operator $\check{\Gamma}(\hat{j}_{R})$ by 
\begin{equation}
\check{\Gamma}(\hat{j}_{R}):=U_{[L^{2}(\mathbb{R}^{d})],[L^{2}(\mathbb{R}^{d})]}\Gamma_{\textrm{b}}(\hat{j}_{R}),
\end{equation}
which acts from $\mathscr{H}$ to $\mathscr{H}^{\textrm{e}}$.
\vspace{2mm}
\\
\textbf{Lemma 5.1.} For any $\chi \in C_{0}^{\infty}(\mathbb{R}) $,
\begin{equation*}
\displaystyle\lim_{R\rightarrow\infty}\big\|\chi(H^{\textrm{e}}_{m})\check{\Gamma}(\hat{j}_{R})-\check{\Gamma}(\hat{j}_{R})\chi(H_{m})\big\|=0.
\end{equation*}
\textit{Proof.}
By the Helffer-Sj\"{o}strand formula [7,16],  it is seen that
\begin{equation}\label{int}
\chi(H^{\textrm{e}}_{m})\check{\Gamma}(\hat{j}_{R})-\check{\Gamma}(\hat{j}_{R})\chi(H_{m})=\frac{-i}{2\pi}\displaystyle\int_{\mathbb{C}}\partial_{\overline{z}}\tilde{\chi}(z)(z-H^{\textrm{e}}_{m})^{-1}\big(H_{m}^{\textrm{e}}\check{\Gamma}(\hat{j}_{R})-\check{\Gamma}(\hat{j}_{R})H_{m}\big)(z-H_{m})^{-1} \textrm{d}z\textrm{d}\overline{z},  
\end{equation}
where $\tilde{\chi}$ is an \textit{almost analytic extension} of $\chi$ and $\partial_{\overline{z}}=\frac{1}{2}(\partial_{x}+i\partial_{y})\hspace{3mm}(z=x+iy)$. Let us estimate the integrand on the left hand side of (\ref{int}). It follows that
\begin{equation*}
   \begin{aligned}
&(z-H^{\textrm{e}}_{m})^{-1}\big(H_{m}^{\textrm{e}}\check{\Gamma}(\hat{j}_{R})-\check{\Gamma}(\hat{j}_{R})H_{m}\big)(z-H_{m})^{-1}\\
&=(z-H^{\textrm{e}}_{m})^{-1}(N_{0}+N_{\infty}+1)(N_{0}+ N_{\infty}+1)^{-1}\big(H_{m}^{\textrm{e}}\check{\Gamma}(\hat{j}_{R})-\check{\Gamma}(\hat{j}_{R})H_{m}\big)(N_{\textrm{b}}+1)^{-1}(N_{\textrm{b}}+1)(z-H_{m})^{-1}.
   \end{aligned}
\end{equation*}
Where, $N_{0}:=N_{\textrm{b}}\otimes1_{\mathscr{H}}$ and $N_{\infty}:= 1_{\mathscr{H}}\otimes N_{\textrm{b}}$. It is easy to see that $(z-H^{\textrm{e}}_{m})^{-1}(N_{0}+N_{\infty}+1)$ is a bounded operator on $D(N_{0}+N_{\infty})$ with operator norm 
\begin{equation*}
\big\|(z-H^{\textrm{e}}_{m})^{-1}(N_{0}+N_{\infty}+1)\big\|\le C\Big(1+(1+|z|)|\text{Im}\hspace{1mm} z|^{-1}\Big),
\end{equation*} 
where $C>0$ is a constant independent of $z$ and we used the fact that $N_{\textrm{b}}$ is $\textrm{d}\Gamma_{\textrm{b}}([\omega_{m}])$ -bounded and the fact that if a linear operator $S$ is bounded, so is $S^{\ast}$. Similarly one can show that $(N_{\textrm{b}}+1)(z-H_{m})^{-1}$ is also a bounded operator with operator norm  
\begin{equation*}
\big\|(N_{\textrm{b}}+1)(z-H_{m})^{-1}\big\|\le D\Big(1+(1+|z|)|\text{Im}\hspace{1mm} z|^{-1}\Big),
\end{equation*}
where $D>0$ is a constant independent of $z$.
Thus we have
\begin{equation*}
 \begin{aligned}
\big\|(z-H^{\textrm{e}}_{m})^{-1}&\big(H_{m}^{\textrm{e}}\check{\Gamma}(\hat{j}_{R})-\check{\Gamma}(\hat{j}_{R})H_{m}\big)(z-H_{m})^{-1}\big\|\\
&\le CD\Big(1+(1+|z|)|\text{Im}\hspace{1mm} z|^{-1}\Big)^{2}\big\|(N_{0}+ N_{\infty}+1)^{-1}\big(H_{m}^{\textrm{e}}\check{\Gamma}(\hat{j}_{R})-\check{\Gamma}(\hat{j}_{R})H_{m}\big)(N_{\textrm{b}}+1)^{-1}\big\|.
   \end{aligned}
\end{equation*}
By the property of $\tilde{\chi}$, it suffices show that 
\begin{equation}\label{norm}
\displaystyle\lim_{R\rightarrow\infty}\big\|(N_{0}+ N_{\infty}+1)^{-1}\big(H_{m}^{\textrm{e}}\check{\Gamma}(\hat{j}_{R})-\check{\Gamma}(\hat{j}_{R})H_{m}\big)(N_{\textrm{b}}+1)^{-1}\big\|=0.
\end{equation}
We have
\begin{equation}\label{hs}
\big(H_{m}^{\textrm{e}}\check{\Gamma}(\hat{j}_{R})-\check{\Gamma}(\hat{j}_{R})H_{m}\big) = \Big\{H^{\textrm{e}}_{0,m}\check{\Gamma}(\hat{j}_{R})-\check{\Gamma}(\hat{j}_{R})\textrm{d}\Gamma_{\textrm{b}}([\omega_{m}])\Big\} + \Big\{(\mu H_{1}+\lambda H_{2})\otimes1_{\mathscr{H}}\check{\Gamma}(\hat{j}_{R})-\check{\Gamma}(\hat{j}_{R})(\mu H_{1}+\lambda H_{2})\Big\}.
\end{equation}
Let us estimate the first term on the right hand side of (\ref{hs}). For any $\Psi\in\mathscr{F}_{\textrm{b,fin}}([C_{0}^{\infty}(\mathbb{R}^{d})])$, it is seen that
\begin{equation}\label{ineq}
  \begin{aligned}
&\hspace{3mm}\big\|\big(H^{\textrm{e}}_{0,m}\check{\Gamma}(\hat{j}_{R})-\check{\Gamma}(\hat{j}_{R})\textrm{d}\Gamma_{\textrm{b}}([\omega_{m}])\big)(N_{\textrm{b}}+1)^{-1}\Psi\big\|^{2}
\\
&= \big\|\textrm{d}\Gamma_{\textrm{b}}(\hat{j}_{R}, \oplus^{2}[\omega_{m}]\hat{j}_{R}-\hat{j}_{R}[\omega_{m}])(N_{\textrm{b}}+1)^{-1}\Psi\big\|^{2}
\\
&\le 4\big(\big\|[\omega_{m}, \hat{j}_{0,R}]\big\|+\big\|[\omega_{m}, \hat{j}_{\infty,R}]\big\|\big)^{2}\big\|\Psi\big\|^{2} \\
&\le \frac{C^{2}}{R^{2}}\big\|\Psi\big\|^{2},
\end{aligned}
\end{equation}
where $C>0$ is a constant and
\begin{equation*}
\textrm{d}\Gamma_{\textrm{b}}(\hat{j}_{R}, \oplus^{2}[\omega_{m}]\hat{j}_{R}-\hat{j}_{R}[\omega_{m}]):=0\oplus\bigoplus_{n=1}^{\infty}\displaystyle\sum_{l=1}^{n}\hat{j}_{R}\otimes\dots\otimes\hat{j}_{R}\otimes\underbrace{\big(\oplus^{2}[\omega_{m}]\hat{j}_{R}-\hat{j}_{R}[\omega_{m}]\big)}_{l\text{-th}}\otimes\hat{j}_{R}\otimes\dots\otimes\hat{j}_{R}.
\end{equation*}
 To derive the last inequality of (\ref{ineq}), let us estimate $\|[\omega_{m}, \hat{j}_{0,R}]\|$ and $\|[\omega_{m}, \hat{j}_{\infty,R}]\|$. For any $f\in L^{1}(\mathbb{R}^{d})\cap L^{2}(\mathbb{R}^{d})$ we see that
\begin{equation*}
(\hat{j}_{0,R}f)(k)
=(2\pi)^{-d/2}\displaystyle\int_{\mathbb{R}^{d}}(\mathcal{F}j_{0})(u)f(k+u/R) \textrm{d}u,
\end{equation*}  
where $\mathcal{F}$ denotes the Fourier transform on $L^{2}(\mathbb{R}^{d})$. We define a positive function $\langle\cdot\rangle$ by 
\begin{equation*}
\langle k\rangle:=\sqrt{k^{2}+1},\hspace{3mm}k\in\mathbb{R}^{d}.
\end{equation*}
For any $g\in C_{0}^{\infty}(\mathbb{R}^{d})$, it is seen that
\begin{equation*}
   \begin{aligned}
 \big\|[\hat{j}_{0}, \tilde{\omega}_{m}]g\big\|^{2}&=\displaystyle\int_{\mathbb{R}^{d}}\big|(\hat{j}_{0,R}\omega_{m}g)(k)-\omega_{m}(k)(\hat{j}_{0,R}g)(k)\big|^{2}\textrm{d}k
\\
&\le(2\pi)^{-d}\displaystyle\int_{\mathbb{R}^{d}}\Big(\displaystyle\int_{\mathbb{R}^{d}}|(\mathcal{F}j_{0})(u)||\omega_{m}(k+u/R)-\omega_{m}(k)||g(k+u/R)|\textrm{d}u\Big)^{2}\textrm{d}k
\\
&\le\frac{1}{(2\pi)^{d}R^{2}}\displaystyle\int_{\mathbb{R}^{d}}\Big(\displaystyle\int_{\mathbb{R}^{d}}|(\mathcal{F}j_{0})(u)||u||g(k+u/R)|\textrm{d}u\Big)^{2}\textrm{d}k
\\
&\le\frac{1}{(2\pi)^{d}R^{2}}\displaystyle\int_{\mathbb{R}^{d}}\Big(\displaystyle\int_{\mathbb{R}^{d}}|(\mathcal{F}j_{0})(u)|\langle u\rangle|g(k+u/R)| \textrm{d}u\Big)^{2} \textrm{d}k
\\
&\le\frac{1}{(2\pi)^{d}R^{2}}\big\|(\mathcal{F}j_{0})\langle\cdot\rangle^{d+1}\big\|^{2}\big\|\langle\cdot\rangle^{-d}\big\|^{2}\big\|g\big\|^{2}.
 \end{aligned}
\end{equation*}
Note that 
\begin{equation*}
\big\|[\hat{j}_{\infty,R},\omega_{m}]\big\|=\big\|[1-\hat{j}_{\infty,R},\omega_{m}]\big\|,
\end{equation*}
and $1-j_{\infty,0}\in C_{0}^{\infty}(\mathbb{R}^{d})$, $\big\|[\hat{j}_{\infty,R},\omega_{m}]\big\|$ is also estimated by the same manner. Hence, we have
\begin{equation*}
\displaystyle\lim_{R\rightarrow\infty}\big\|(N_{0}+N_{\infty}+1)^{-1}(H^{\textrm{e}}_{0,m}\check{\Gamma}(\hat{j}_{R})-\check{\Gamma}(\hat{j}_{R})\textrm{d}\Gamma_{\textrm{b}}([\omega_{m}])\big)(N_{\textrm{b}}+1)^{-1}\big\|=0.
\end{equation*}
Next we estimate $(N_{0}+N_{\infty}+1)^{-1}(H_{2}\otimes1_{\mathscr{H}}\check{\Gamma}(\hat{j}_{R})-\check{\Gamma}(\hat{j}_{R}))H_{2})(N_{\textrm{b}}+1)^{-1}$. For any $\Psi\in \mathscr{F}_{\textrm{b,fin}}([C_{0}^{\infty}(\mathbb{R}^{d})])$, it is seen that 
\begin{equation}\label{3}
\begin{aligned}
&(N_{0}+N_{\infty}+1)^{-1}(H_{2}\otimes1_{\mathscr{H}}\check{\Gamma}(\hat{j}_{R})-\check{\Gamma}(\hat{j}_{R}))H_{2})(N_{\textrm{b}}+1)^{-1}\Psi
\\
=&\displaystyle\int_{\mathbb{R}^{d}}\chi_{\textrm{sp}}(x)(N_{0}+N_{\infty}+1)^{-1}\Big((\phi(f_{x})^{\ast}\phi(f_{x}))^{2}\otimes1_{\mathscr{H}}\check{\Gamma}(\hat{j}_{R})-\check{\Gamma}(\hat{j}_{R})((\phi(f_{x})^{\ast}\phi(f_{x}))^{2}\Big)(N_{\textrm{b}}+1)^{-1}\Psi \textrm{d}x.
\end{aligned}
\end{equation}
The integrand on the right hand side of (\ref{3}) is decomposed as follows:
\begin{equation*}
\begin{aligned}
&\hspace{5mm}(N_{0}+N_{\infty}+1)^{-1}\Big((\phi(f_{x})^{\ast}\phi(f_{x}))^{2}\otimes1_{\mathscr{H}}\check{\Gamma}(\hat{j}_{R})-\check{\Gamma}(\hat{j}_{R})((\phi(f_{x})^{\ast}\phi(f_{x}))^{2}\Big)(N_{\textrm{b}}+1)^{-1}\Psi
\\
&=(N_{0}+N_{\infty}+1)^{-1}\big(\phi(f_{x})^{\ast}\phi(f_{x})\phi(f_{x})^{\ast}\otimes1_{\mathscr{H}}\big)\Big(\phi(f_{x})\otimes1_{\mathscr{H}}\check{\Gamma}(\hat{j}_{R})-\check{\Gamma}(\hat{j}_{R})\phi(f_{x})\Big)(N_{\textrm{b}}+1)^{-1}\Psi  \\
&\hspace{3mm}
+(N_{0}+N_{\infty}+1)^{-1}\big(\phi(f_{x})^{\ast}\phi(f_{x}))\otimes1_{\mathscr{H}}\big)\Big(\phi(f_{x})^{\ast}\otimes1_{\mathscr{H}}\check{\Gamma}(\hat{j}_{R})-\check{\Gamma}(\hat{j}_{R})\phi(f_{x})^{\ast}\Big)\phi(f_{x})(N_{\textrm{b}}+1)^{-1}\Psi   \\
&\hspace{3mm}
+(N_{0}+N_{\infty}+1)^{-1}\big(\phi(f_{x})^{\ast}\otimes1_{\mathscr{H}}\big)\Big(\phi(f_{x})\otimes1_{\mathscr{H}}\check{\Gamma}(\hat{j}_{R})-\check{\Gamma}(\hat{j}_{R})\phi(f_{x})\Big)\phi(f_{x})^{\ast}\phi(f_{x})(N_{\textrm{b}}+1)^{-1}\Psi   \\
&\hspace{3mm}
+(N_{0}+N_{\infty}+1)^{-1}\Big(\phi(f_{x})^{\ast}\otimes1_{\mathscr{H}}\check{\Gamma}(\hat{j}_{R})-\check{\Gamma}(\hat{j}_{R})\phi(f_{x})^{\ast}\Big)\phi(f_{x})\phi(f_{x})^{\ast}\phi(f_{x})(N_{\textrm{b}}+1)^{-1}\Psi     
\\
&=M_{1}\big(\phi(f_{x})^{\ast}\otimes1_{\mathscr{H}}\big)\Big(\phi((1-\hat{j}_{0,R})f_{x})\otimes1_{\mathscr{H}}-1_{\mathscr{H}}\otimes\phi(\hat{j}_{\infty,R}f_{x})\Big)\check{\Gamma}(\hat{j}_{R})(N_{\textrm{b}}+1)^{-1}\Psi  \\
&\hspace{3mm}
+M_{2}\Big(\phi((1-\hat{j}_{0,R})f_{x})^{\ast}\otimes1_{\mathscr{H}}-1_{\mathscr{H}}\otimes\phi(\hat{j}_{\infty,R}f_{x})^{\ast}\Big)\check{\Gamma}(\hat{j}_{R})\phi(f_{x})(N_{\textrm{b}}+1)^{-1}\Psi  \\
&\hspace{3mm}
+M_{3}\check{\Gamma}(\hat{j}_{R})\phi(f_{x})^{\ast}\phi(f_{x})(N_{\textrm{b}}+1)^{-1}\Psi  \\
&\hspace{3mm}
+M_{4}\check{\Gamma}(\hat{j}_{R})\phi(f_{x})^{\ast}\phi(f_{x})(N_{\textrm{b}}+1)^{-1}\Psi,
  \end{aligned}
\end{equation*}
where
\begin{equation*}
\begin{aligned}
M_{1}&:=(N_{0}+N_{\infty}+1)^{-1}\big(\phi(f_{x})^{\ast}\phi(f_{x})\big)\otimes1_{\mathscr{H}},
\\
M_{2}&:=(N_{0}+N_{\infty}+1)^{-1}\big(\phi(f_{x})^{\ast}\phi(f_{x})\big)\otimes1_{\mathscr{H}},
\\
M_{3}&:=(N_{0}+N_{\infty}+1)^{-1}\big(\phi(f_{x})^{\ast}\otimes1_{\mathscr{H}}\big)\Big(\phi((1-\hat{j}_{0,R})f_{x})\otimes1_{\mathscr{H}}-1_{\mathscr{H}}\otimes\phi(\hat{j}_{\infty,R}f_{x})\Big),
\\
M_{4}&:=(N_{0}+N_{\infty}+1)^{-1}\Big(\phi((1-\hat{j}_{0,R})f_{x})^{\ast}\otimes1_{\mathscr{H}}-1_{\mathscr{H}}\otimes\phi(\hat{j}_{\infty,R}f_{x})^{\ast}\Big)\Big(\phi(\hat{j}_{0,R}f_{x})\otimes1_{\mathscr{H}}+1_{\mathscr{H}}\otimes\phi(\hat{j}_{\infty,R}f_{x})\Big).
\end{aligned}
\end{equation*}
Since $M_{j}\hspace{2mm}(j=1,2,3,4)$ are bounded operator on $D(N_{0})\cap D(N_{\infty})$ respectively and
\begin{equation*}
(N_{\textrm{b}}+1)^{-1}\Psi\in\mathscr{F}_{\textrm{b,fin}}([C_{0}^{\infty}(\mathbb{R}^{d})]),\hspace{3mm} \check{\Gamma}(\hat{j}_{R})\Psi\in \mathscr{F}_{\textrm{b,fin}}([L^{2}(\mathbb{R}^{d})])\hat{\otimes} \mathscr{F}_{\textrm{b,fin}}([L^{2}(\mathbb{R}^{d})]),
\end{equation*} 
it follows that
\begin{equation*}
\begin{aligned}
&\hspace{5mm}\big\|(N_{0}+N_{\infty}+1)^{-1}((\phi(f_{x})^{\ast}\phi(f_{x}))^{2}\otimes1_{\mathscr{H}}\check{\Gamma}(\hat{j}_{R})-\check{\Gamma}(\hat{j}_{R})((\phi(f_{x})^{\ast}\phi(f_{x}))^{2})(N_{\textrm{b}}+1)^{-1}\Psi\big\|
\\
&\le \big\|M_{1}\big\|\big\|\big(\phi(f_{x})^{\ast}\otimes1_{\mathscr{H}}\big)\big(\phi((1-\hat{j}_{0,R})f_{x})\otimes1_{\mathscr{H}}+1_{\mathscr{H}}\otimes\phi(\hat{j}_{\infty,R}f_{x})\big)\check{\Gamma}(\hat{j}_{R})(N_{\textrm{b}}+1)^{-1}\Psi\big\|  \\
&\hspace{2mm}+\big\|M_{2}\big\| \big\|\big(\phi((1-\hat{j}_{0,R})f_{x})^{\ast}\otimes1_{\mathscr{H}}+1_{\mathscr{H}}\otimes\phi(\hat{j}_{\infty,R}f_{x})^{\ast}\big)\big(\phi(\hat{j}_{0,R}f_{x})\otimes1_{\mathscr{H}}+1_{\mathscr{H}}\otimes\phi(\hat{j}_{\infty,R}f_{x})\big)\check{\Gamma}(\hat{j}_{R})(N_{\textrm{b}}+1)^{-1}\Psi\big\|  \\
&\hspace{2mm}
+\big\|M_{3}\big\|\big(\big\|(1-\hat{j}_{0,R})f_{x}\big\|+\big\|\hat{j}_{\infty,R}f_{x}\big\|\big)\big\|\check{\Gamma}(\hat{j}_{R})\phi(f_{x})^{\ast}\phi(f_{x})(N_{\textrm{b}}+1)^{-1}\Psi \big\|  \\
&\hspace{2mm}
+\big\|M_{4}\big\|\big(\big\|(1-\hat{j}_{0,R})f_{x}\big\|+\big\|\hat{j}_{\infty,R}f_{x}\big\|\big)\big\|\phi(f_{x})^{\ast}\phi(f_{x})(N_{\textrm{b}}+1)^{-1}\Psi \big\|  \\
&\le D\big(\big\|(1-\hat{j}_{0,R})f_{x}\big\|+\big\|\hat{j}_{\infty,R}f_{x}\big\|\big)\big\|\Psi\big\|.
   \end{aligned}
\end{equation*}
Since $\mathscr{F}_{\textrm{b,fin}}([C_{0}^{\infty}(\mathbb{R}^{d})])$ is dense in $\mathscr{H}$,  it follows from an application of the extension theorem of bounded operators that 
\begin{equation*}
\big\|(N_{0}+N_{\infty}+1)^{-1}(H_{2}\otimes1_{\mathscr{H}}\check{\Gamma}(\hat{j}_{R})-\check{\Gamma}(\hat{j}_{R})H_{2})(N_{\textrm{b}}+1)^{-1}\big\| 
\le D\displaystyle\int_{\mathbb{R}^{d}}\chi_{\textrm{sp}}(x) \big(\big\|(1-\hat{j}_{0,R})f_{x}\big\|+\big\|\hat{j}_{\infty,R}f_{x}\big\|\big)\textrm{d}x.
\end{equation*}
Since $\lim_{R\rightarrow\infty}\big\|(1-\hat{j}_{0,R})f_{x}\big\| ,\lim_{R\rightarrow\infty}\big\|\hat{j}_{\infty,R}f_{x}\big\|=0$, we have
\begin{equation*}
\displaystyle\lim_{R\rightarrow\infty}\big\||(N_{0}+N_{\infty}+1)^{-1}\big(H_{2}\otimes1_{\mathscr{H}}\check{\Gamma}(\hat{j}_{R})-\check{\Gamma}(\hat{j}_{R}))H_{2}\big)(N_{\textrm{b}}+1)^{-1}\big\|=0,
\end{equation*}
 by using the Lebesgue dominated convergence theorem.
Similarly we can show that $\lim_{R\rightarrow\infty}\big\||(N_{0}+N_{\infty}+1)^{-1}\big(H_{1}\otimes1_{\mathscr{H}}\check{\Gamma}(\hat{j}_{R})-\check{\Gamma}(\hat{j}_{R})H_{1}\big)(N_{\textrm{b}}+1)^{-1}\big\|=0$. Therefore we obtain the desired result.\hspace{32mm}$\square$
\vspace{2mm}
\\
\textbf{Lemma 5.2.}
 For any $\chi\in C_{0}^{\infty}(\mathbb{R})$ such that supp\hspace{1mm}$\chi \subset (-\infty , E_{0}(H_{m})+m)$, $\chi(H_{m})$ is a compact operator. Especially, $H_{m}$ has a ground state.
\begin{proof}
Let $E_{N_{\textrm{b}}}$ be the spectral measure of $N_{\textrm{b}}$. For any $n\in\mathbb{N}$, it follows that
\begin{equation*}
E_{N_{\textrm{b}}}(\{n\})\Gamma_{\textrm{b}}([\hat{j}_{0,R}^{2}])\chi(H_{m})=E_{N_{\textrm{b}}}(\{n\})\Gamma_{\textrm{b}}([\hat{j}_{0,R}^{2}])(\textrm{d}\Gamma_{\textrm{b}}([\omega_{m}])+1)^{-1}(\textrm{d}\Gamma_{\textrm{b}}([\omega_{m}])+1)\chi(H_{m})= J_{1}J_{2},
\end{equation*}
where
\begin{equation*}
\begin{aligned}
J_{1}&:=E_{N_{\textrm{b}}}(\{n\})\Gamma_{\textrm{b}}([\hat{j}_{0,R}^{2}])(\textrm{d}\Gamma_{\textrm{b}}([\omega_{m}])+1)^{-1},
\\
J_{2}&:=(\textrm{d}\Gamma_{\textrm{b}}([\omega_{m}])+1)\chi(H_{m}).
\end{aligned}
\end{equation*}
Since $J_{1}$ is compact (see [8, Lemma 4.2]) and $J_{2}$ is bounded, $E_{N_{\textrm{b}}}(\{n\})\Gamma_{\textrm{b}}([\hat{j}_{0,R}^{2}])\chi(H_{m})$ is compact. Note that  
\begin{equation*}
\big\|\Gamma_{\textrm{b}}([\hat{j}_{0,R}^{2}])\chi(H_{m})-\displaystyle\sum_{n=1}^{N}E_{N_{\textrm{b}}}(\{n\})\Gamma_{\textrm{b}}([\hat{j}_{0,R}^{2}])\chi(H_{m})\big\|\le \frac{1}{N+1}\big\|\Gamma_{\textrm{b}}([\hat{j}_{0,R}^{2}])(N_{\textrm{b}}+1)\chi(H_{m})\big\|\rightarrow0,\hspace{2mm}N\rightarrow\infty.
\end{equation*}
Thus $\Gamma_{\textrm{b}}([j_{0,R}^{2}])\chi(H_{m})$ is compact.
Next we show that $\chi(H_{m})$ is compact. Since supp\hspace{1mm}$\chi\subset (-\infty, E_{0}(H_{m})+m)$, it follows that
\begin{equation}
\chi(H_{m}^{\textrm{e}})=(1_{\mathscr{H}}\otimes P_{0})\chi(H_{m}^{\textrm{e}}),
\end{equation}
where $P_{0}$ is the orthogonal projection onto the subspace $\{z\Omega: z\in\mathbb{C}\}$. Furthermore, the following property also holds:
\begin{equation*}
\check{\Gamma}(\hat{j}_{R})^{\ast}\check{\Gamma}(\hat{j}_{R})=1_{\mathscr{H}^{\textrm{e}}},\hspace{5mm}\check{\Gamma}(\hat{j}_{R})^{\ast}(1_{\mathscr{H}}\otimes P_{0})\check{\Gamma}(\hat{j}_{R})=\Gamma_{\textrm{b}}([\hat{j}_{0,R}^{2}]).
\end{equation*}
By applying Lemma 5.1, we have
\begin{equation*}
\begin{aligned}
\chi(H_{m})&=\check{\Gamma}(\hat{j}_{R})^{\ast}\check{\Gamma}(\hat{j}_{R})\chi(H_{m})
\\
&=\check{\Gamma}(\hat{j}_{R})^{\ast}\chi(H_{m}^{\textrm{e}})\check{\Gamma}(\hat{j}_{R})+o(R^{0})
\\
&=\check{\Gamma}(\hat{j}_{R})^{\ast}(1_{\mathscr{H}}\otimes P_{0})\chi(H_{m}^{\textrm{e}})\check{\Gamma}(\hat{j}_{R})+o(R^{0})
\\
&=\check{\Gamma}(\hat{j}_{R})^{\ast}(1_{\mathscr{H}}\otimes P_{0})\check{\Gamma}(\hat{j}_{R})\chi(H_{m})+o(R^{0})
\\
&=\Gamma_{\textrm{b}}([\hat{j}_{0,R}^{2}])\chi(H_{m})+o(R^{0}),
\end{aligned}
\end{equation*}
where $o(R^{0})$ denotes a bounded operator tending to 0 as $R\rightarrow\infty$ in operator norm topology. Thus $\chi(H_{m})$ is compact. By applying a general theorem [21, Theorem XIII-77], one sees that $\sigma(H_{m})\cap(-\infty, E_{0}(H_{m})+m)$ is purely discrete. In particular, $E_{0}(H_{m})$ is an eigenvalue of $H_{m}$.
\end{proof}
For $m>0$, let $\Phi_{m}$ be a ground state of $H_{m}$ with $\|\Phi_{m}\|=1$.
\vspace{2mm}
\\
\textbf{Lemma 5.3.} $H_{m}\rightarrow H$ (as $m\rightarrow0$) \textit{in the strong resolvent sense. Especially,} $E_{0}(H_{m})\rightarrow E_{0}(H)\hspace{2mm}(\text{as}\hspace{2mm}m\rightarrow0$).
\begin{proof}
 For any $\Psi\in\mathscr{F}_{\textrm{b,fin}}([C_{0}^{\infty}(\mathbb{R}^{d})])$, we have $H_{m}\Psi\rightarrow H\Psi\hspace{2mm}(\text{as}\hspace{2mm}m\rightarrow0)$ by direct calculation. This fact implies the strong resolvent convergence [20, Theorem VIII 25 (a)]. The strong resolvent convergence implies that $\limsup_{m\rightarrow0}E_{0}(H_{m})\le E_{0}(H)$. For any $m>0$, we have
\begin{equation}\label{4}
E_{0}(H_{m})=\langle\Phi_{m}, H_{m}\Phi_{m}\rangle \ge \langle \Phi_{m},H\Phi_{m}\rangle\ge E_{0}(H).
\end{equation}
By taking $\liminf$ on the both side of (\ref{4}), we obtain the desired result.
\end{proof}
For each $n\in\mathbb{N}$, we denote the permutation group of $\{1,\dots, n\}$ by $\mathcal{S}_{n}$. We can identify $\mathscr{H}$ as follows:
 \begin{equation*}
\mathscr{H}=\displaystyle\bigoplus_{n,n'=0}^{\infty}L^{2}_{\textrm{sym}}(\mathbb{R}^{dn}\times\mathbb{R}^{dn'}),
\end{equation*}
where
\begin{equation*}
L^{2}_{\textrm{sym}}(\mathbb{R}^{dn}):=\big\{f\in L^{2}(\mathbb{R}^{dn}): f(k_{\pi(1)},\cdots,k_{\pi(n)})=f(k_{1},\cdots,k_{n}) \hspace{2mm}\text{for a,e, $k_{1}, \cdots , k_{n} \in\mathbb{R}^{d}$ and $\pi\in \mathcal{S}_{n}$}\big\},
\end{equation*}
\begin{equation*}
L^{2}_{\textrm{sym}}(\mathbb{R}^{dn}\times\mathbb{R}^{0}):=L^{2}_{\textrm{sym}}(\mathbb{R}^{dn}),\hspace{2mm}L^{2}_{\textrm{sym}}(\mathbb{R}^{0}\times\mathbb{R}^{dn'}):=L^{2}_{\textrm{sym}}(\mathbb{R}^{dn'}),\hspace{2mm} L_{\textrm{sym}}^{2}(\mathbb{R}^{0}\times\mathbb{R}^{0}):=\mathbb{C},
\end{equation*}
\begin{equation*}
\begin{aligned}
L_{\textrm{sym}}^{2}(\mathbb{R}^{dn}\times\mathbb{R}^{dn'}):&=\{f\in L^{2}(\mathbb{R}^{d(n+n')}): \text{for a.e.}\hspace{2mm} k_{1},\dots k_{n}, l_{1},\dots l_{n'} \in\mathbb{R}^{d}, \sigma\in\mathcal{S}_{n}, \tau\in\mathcal{S}_{n'},
\\
&\hspace{30mm}f(k_{\sigma(1)},\dots,k_{\sigma(n)}:l_{\tau(1)},\dots,l_{\tau(n')})=f(k_{1},\dots,k_{n}:l_{1},\dots,l_{n'})\}.
\end{aligned}
\end{equation*}
For $k\in\mathbb{R}^{d}$, linear operator $a_{+}(k)$ and $a_{-}(k)$ act on $\mathscr{H}$ are defined as follows:
\begin{equation*}
(a_{+}(k)\Psi)^{(n,n')}(k_{1}, \dots, k_{n}:l_{1}, \dots, l_{n'} ):=\sqrt{n+1}\Psi^{(n+1,n')}(k, k_{1}, \dots, k_{n}: l_{1}, \dots, l_{n'})\hspace{2mm}a.e.,
\end{equation*}
\begin{equation*}
(a_{-}(k)\Psi)^{(n,n')}(k_{1}, \dots, k_{n}:l_{1}, \dots, l_{n'} ):=\sqrt{n'+1}\Psi^{(n,n'+1)}(k_{1}, \dots, k_{n}:k, l_{1}, \dots, l_{n'})\hspace{2mm}a.e..
\end{equation*}
$a_{+}(\cdot)$ and $a_{-}(\cdot)$ are called the annihilation kernel of particle and anti-particle respectively. For each $u\in L^{2}(\mathbb{R}^{d})$, $a_{+}(u)$ and $a_{-}(u)$ are represented  by using the annihilation kernel as follows:
\begin{equation}\label{ann}
a_{\pm}(u)=\displaystyle\int_{\mathbb{R}^{d}}u(k)^{\ast}a_{\pm}(k)\textrm{d}k,
\end{equation} 
where the integrals on the right hand side of (\ref{ann}) are taken in the sense of $\mathscr{H}$-valued strong Bochner integral. For $k\in\mathbb{R}^{d}$, let us introduce the following operators:
\begin{equation*}
\begin{aligned}
S_{1}(k)&:=\displaystyle\int_{\mathbb{R}^{d}}\chi_{\textrm{sp}}(x)e^{-ikx}\phi(f_{x})\textrm{d}x,
\hspace{5mm}
S_{2}(k):=\displaystyle\int_{\mathbb{R}^{d}}\chi_{\textrm{sp}}(x)e^{-ikx}\phi(f_{x})\phi(f_{x})^{\ast}\phi(f_{x})\textrm{d}x,
\\
L_{1}(k)&:=\displaystyle\int_{\mathbb{R}^{d}}\chi_{\textrm{sp}}(x)e^{-ikx}\phi(f_{x})^{\ast}\textrm{d}x,
\hspace{4mm}
L_{2}(k):=\displaystyle\int_{\mathbb{R}^{d}}\chi_{\textrm{sp}}(x)e^{-ikx}\phi(f_{x})^{\ast}\phi(f_{x})\phi(f_{x})^{\ast}\textrm{d}x.
\end{aligned}
\end{equation*}
Note that these operator are also taken in the sense of $\mathscr{H}$-valued strong Bochner integral.
\vspace{2mm}
\\
\textbf{Lemma 5.4.}
\textit{For $k\neq0$, we have}
\begin{equation}\label{ptf}
\begin{aligned}
a_{+}(k)\Phi_{m}&=\frac{\varphi(k)}{\sqrt{2\omega(k)}}\big(E_{0}(H_{m})-H_{m}-\omega_{m}(k)\big)^{-1}\big(\mu S_{1}(k)+2\lambda \overline{S_{2}(k)}\big)\Phi_{m},
\\
a_{-}(k)\Phi_{m}&=\frac{\varphi(k)}{\sqrt{2\omega(k)}}\big(E_{0}(H_{m})-H_{m}-\omega_{m}(k)\big)^{-1}\big(\mu L_{1}(k)+2\lambda \overline{L_{2}(k)}\big)\Phi_{m}.
\end{aligned}
\end{equation}
\textit{Proof.}
Here, we only prove the equation about $a_{+}(k)\Phi_{m}$. The case of $a_{-}(k)\Phi_{m}$ is proven similarly. Let $\Theta\in\mathscr{F}_{\textrm{b,fin}}([C_{0}^{\infty}(\mathbb{R}^{d})])$ and $g\in C_{0}^{\infty}(\mathbb{R}^{d})$. Since $\Phi_{m}\in \text{Ker}(H_{m}-E_{0}(H_{m}))$, we have
\begin{equation}\label{aiu}
\begin{aligned}
&\hspace{4mm}\langle(H_{m}-E_{0}(H_{m}))\Theta, a_{+}(g)\Phi_{m}\rangle
\\
&=\langle [a_{+}(g)^{\dag},H_{m}-E_{0}(H_{m})]\Theta, \Phi_{m}\rangle
\\
&=-\langle a_{+}(\omega_{m}g)^{\dag}\Theta,\Phi_{m}\rangle-\frac{1}{\sqrt{2}}\displaystyle\int_{\mathbb{R}^{d}}\chi_{\textrm{sp}}(x)\langle g,f_{x}\rangle \langle \big(\mu\phi(f_{x})^{\ast}+2\lambda\phi(f_{x})^{\ast}\phi(f_{x})\phi(f_{x})^{\ast}\big)\Theta, \Phi_{m}\rangle \textrm{d}x
\\
&=-\langle a_{+}(\omega_{m}g)^{\dag}\Theta,\Phi_{m}\rangle-\displaystyle\int_{\mathbb{R}^{d}}g(k)^{\ast}\frac{\varphi(k)}{\sqrt{2\omega(k)}}\textrm{d}k\displaystyle\int_{\mathbb{R}^{d}}\chi_{\textrm{sp}}(x)e^{-ikx}\langle \big(\mu\phi(f_{x})^{\ast}+2\lambda\phi(f_{x})^{\ast}\phi(f_{x})\phi(f_{x})^{\ast}\big)\Theta, \Phi_{m}\rangle \textrm{d}x.
\end{aligned}
\end{equation}
Here to get the last equality of (\ref{aiu}), we used Fubini's theorem. By using (\ref{ann}), we have
\begin{equation*}
\begin{aligned}
&\hspace{6mm}\displaystyle\int_{\mathbb{R}^{d}}g(k)^{\ast}\langle (E_{0}(H_{m})-H_{m}-\omega_{m}(k))\Theta, a_{+}(k)\Phi_{m}\rangle \textrm{d}k
\\
&=\displaystyle\int_{\mathbb{R}^{d}}g(k)^{\ast}\frac{\varphi(k)}{\sqrt{2\omega(k)}}\textrm{d}k\displaystyle\int_{\mathbb{R}^{d}}\chi_{\textrm{sp}}(x)e^{-ikx}\langle \big(\mu\phi(f_{x})^{\ast}+2\lambda\phi(f_{x})^{\ast}\phi(f_{x})\phi(f_{x})^{\ast}\big)\Theta, \Phi_{m}\rangle \textrm{d}x,
\end{aligned}
\end{equation*}
Since $g\in C_{0}^{\infty}(\mathbb{R}^{d})$ is arbitrary, we obtain
\begin{equation*}
\langle (E_{0}(H_{m})-H_{m}-\omega_{m}(k))\Theta, a_{+}(k)\Phi_{m}\rangle=\frac{\varphi(k)}{\sqrt{2\omega(k)}}\displaystyle\int_{\mathbb{R}^{d}}\chi_{\textrm{sp}}(x)e^{-ikx}\langle\big(\mu\phi(f_{x})^{\ast}+2\lambda \phi(f_{x})^{\ast}\phi(f_{x})\phi(f_{x})^{\ast}\big)\Theta, \Phi_{m}\rangle \textrm{d}x.
\end{equation*}
Since $\Phi_{m}\in D(H_{m})$, there exists a sequence $\{\Phi_{m}^{j}\}_{j=1}^{\infty}\subset \mathscr{F}_{\textrm{b,fin}}([C_{0}^{\infty}(\mathbb{R}^{d})])$ such that $\Phi_{m}^{j}\rightarrow \Phi_{m}$, $H_{m}\Phi_{m}^{j}\rightarrow H_{m}\Phi_{m}$ $(j\rightarrow 0)$. Therefore we have
\begin{equation*}
\langle (E_{0}(H_{m})-H_{m}-\omega_{m}(k))\Theta, a_{+}(k)\Phi_{m}\rangle =\frac{\varphi(k)}{\sqrt{2\omega(k)}}\langle \Theta, \mu S_{1}(k)\Phi_{m}\rangle +\frac{\varphi(k)}{\sqrt{2\omega(k)}}\displaystyle\lim_{j\rightarrow\infty}\langle \Theta, 2\lambda S_{2}(k)\Phi_{m}^{j}\rangle,
\end{equation*}
where we have used the $H_{m}$-boundedness of $S_{1}(k)$.
We show that for any $k\in\mathbb{R}^{d}$, $S_{2}(k)$ is $H_{m}$-bounded on $\mathscr{F}_{\textrm{b,fin}}([C_{0}^{\infty}(\mathbb{R}^{d})])$. For $\Psi\in\mathscr{F}_{\textrm{b,fin}}([C_{0}^{\infty}(\mathbb{R}^{d})])$, It follows that
\begin{equation*}
\begin{aligned}
\big\|S_{2}(k)\Psi\big\|^{2}&\le\displaystyle\int_{\mathbb{R}^{d}\times\mathbb{R}^{d}}\chi_{\textrm{sp}}(x)\chi_{\textrm{sp}}(y) \big|\langle \phi(f_{x})^{\ast}\phi(f_{x})\phi(f_{y})^{\ast}\phi(f_{y})\Psi, \phi(f_{y}) \phi(f_{x})^{\ast}\Psi\rangle\big| \textrm{d}x\textrm{d}y
\\
&\le \frac{1}{2}\displaystyle\int_{\mathbb{R}^{d}\times\mathbb{R}^{d}}\chi_{\textrm{sp}}(x)\chi_{\textrm{sp}}(y)\langle \big(\phi(f_{y})^{\ast}\phi(f_{y})\big)^{2}\Psi, \big(\phi(f_{x})^{\ast}\phi(f_{x})\big)^{2}\Psi\rangle \textrm{d}x\textrm{d}y 
\\
&\hspace{30mm}+\frac{1}{2}\displaystyle\int_{\mathbb{R}^{d}\times\mathbb{R}^{d}}\chi_{\textrm{sp}}(x)\chi_{\textrm{sp}}(y)\langle \phi(f_{y})^{\ast}\phi(f_{y})\Psi, \phi(f_{x})^{\ast}\phi(f_{x})\Psi\rangle \textrm{d}x\textrm{d}y
\\
&=\frac{1}{2}\big(\big\|H_{2}\Psi\big\|^{2}+\big\|H_{1}\Psi\big\|^{2}\big),
\end{aligned}
\end{equation*}
Therefore $S_{2}(k)$ is $H_{m}$-bounded by Remark 5.1. This fact implies that $\{S_{2}(k)\Phi_{m}^{j}\}_{j=1}^{\infty}$ is a Cauchy sequence. By the closability of $S_{2}(k)$, we have
\begin{equation*}
\langle (E_{0}(H_{m})-H_{m}-\omega_{m}(k))\Theta, a_{+}(k)\Phi_{m}\rangle =\langle \Theta, \mu S_{1}(k)\Phi_{m}\rangle +\langle \Theta, 2\lambda \overline{S_{2}(k)}\Phi_{m}\rangle.
\end{equation*}
Thus we see that $a_{+}(k)\Phi_{m}\in D((E_{0}(H_{m})-H_{m})-\omega_{m}(k))$ and 
\begin{equation*}
(E_{0}(H_{m})-H_{m}-\omega_{m}(k))a_{+}(k)\Phi_{m}=\frac{\varphi(k)}{\sqrt{2\omega(k)}}\big(\mu S_{1}(k)+2\lambda \overline{S_{2}(k)}\big)\Phi_{m}.
\end{equation*}
Since $E_{0}(H_{m})-H_{m}-\omega_{m}(k)$ has a bounded inverse, the equation about $a_{+}(k)\Phi_{m}$ follows.\hspace{35mm}$\square$
\vspace{2mm}
\\
\textbf{Lemma 5.5.} 
\textit{Suppose that} $\varphi\in D(\omega^{-3/2})$, \textit{then} $\Phi_{m}\in D(N_{\textrm{b}}^{1/2})$ \textit{and} 
\begin{equation*}
\displaystyle\sup_{0<m\le1}\|N_{\textrm{b}}^{1/2}\Phi_{m}\| < \infty.
\end{equation*}
\textit{Proof.}
By Proposition A.3 and Proposition A.5, we see that 
\begin{equation*}
\big\|N_{\textrm{b}}^{1/2}\Phi_{m}\big\|^{2}= \displaystyle\int_{\mathbb{R}^{d}}\big\|a_{+}(k)\Phi_{m}\big\|^{2}\textrm{d}k+\displaystyle\int_{\mathbb{R}^{d}}\big\|a_{-}(k)\Phi_{m}\big\|^{2}\textrm{d}k.
\end{equation*}
Let $0<m\le1$. Note that $S_{1}(k), S_{2}(k), L_{1}(k)$ and $L_{2}(k)$ are $H_{m}$-bounded uniformly in $k$. By Lemma 5.4 and $\big\|(E_{0}(H_{m})-H_{m}-\omega_{m}(k))^{-1}\big\|\le\omega(k)^{-1}$, we have
\begin{equation*}
\begin{aligned}
\big\|N_{\textrm{b}}^{1/2}\Phi_{m}\big\|^{2}&\le 
\displaystyle\int_{\mathbb{R}^{d}}\frac{|\varphi(k)|^{2}}{2\omega(k)}\big\|\big(E_{0}(H_{m})-H_{m}-\omega_{m}(k)\big)^{-1}\big(\mu S_{1}(k)+2\lambda \overline{S_{2}(k)}\big)\Phi_{m}\big\|^{2}\textrm{d}k
\\
&\hspace{10mm}+\displaystyle\int_{\mathbb{R}^{d}}\frac{|\varphi(k)|^{2}}{2\omega(k)}\big\|\big(E_{0}(H_{m})-H_{m}-\omega_{m}(k)\big)^{-1}\big(\mu L_{1}(k)+2\lambda \overline{L_{2}(k)}\big)\Phi_{m}\big\|^{2}\textrm{d}k
\\
&\le C(|\mu|^{2}+4\lambda^{2})\big(\big\|H_{m}\Phi_{m}\big\|^{2}+\big\|\Phi_{m}\big\|^{2}\big)\displaystyle\int_{\mathbb{R}^{d}}\frac{|\varphi(k)|^{2}}{\omega(k)^{3}}\textrm{d}k
\\
&=C(|\mu|^{2}+4\lambda^{2})\big(E_{0}(H_{m})^{2}+1\big)\big\|\omega^{-3/2}\varphi\big\|^{2}_{L^{2}},
\end{aligned}
\end{equation*}
where $C>0$ is a constant independent of $m$. Since $\{E_{0}(H_{m})\}_{0<m\le1}$ is bounded by Lemma 5.3, the desired result follows by taking the supremum.\hspace{117mm}$\square$
\vspace{2mm}
\\
\textbf{Lemma 5.6.}
\textit{Suppose that $\varphi$ is differentiable}, $\varphi\in D(\omega^{-3/2})$, $\partial_{j}\varphi\in D(\omega^{-3/2})$ $(j=1,\dots, d)$ \textit{and} 
\\
$\int_{\mathbb{R}^{d}}(1+|x|^{2})\chi_{\textrm{sp}}(x)\textrm{d}x <\infty$. \textit{Then} $a_{\pm}(\cdot)\Phi_{m}$ \textit{is strong differentiable in} $\mathscr{H}$. \textit{Moreover, for} $k\neq0$, 
\begin{equation*}
\begin{aligned}
(D_{j}a_{+})(k)\Phi_{m}&=\frac{2(\partial_{j}\varphi)(k)\omega(k)^{2}-\varphi(k)k_{j}}{2\sqrt{2}\omega(k)^{5/2}}(E_{0}(H_{m})-H_{m}-\omega_{m}(k))^{-1}(\mu S_{1}(k)+2\lambda\overline{S_{2}(k)})\Phi_{m}
\\
&\hspace{3mm}+\frac{k_{j}\varphi(k)}{\omega_{m}(k)\sqrt{2\omega(k)}}\big(E_{0}(H_{m})-H_{m}-\omega_{m}(k)\big)^{-2}(\mu S_{1}(k)+2\lambda\overline{S_{2}(k)})\Phi_{m}
\\
&\hspace{3mm}-\frac{i\varphi(k)}{\sqrt{2\omega(k)}}\big(E_{0}(H_{m})-H_{m}-\omega_{m}(k)\big)^{-1}\big(\mu S_{1.j}(k)+2\lambda\overline{S_{2.j}(k)}\big)\Phi_{m},
\end{aligned}
\end{equation*}
\begin{equation*}
\begin{aligned}
(D_{j}a_{-})(k)\Phi_{m}&=\frac{2(\partial_{j}\varphi)(k)\omega(k)^{2}-\varphi(k)k_{j}}{2\sqrt{2}\omega(k)^{5/2}}(E_{0}(H_{m})-H_{m}-\omega_{m}(k))^{-1}(\mu L_{1}(k)+2\lambda\overline{L_{2}(k)})\Phi_{m}
\\
&\hspace{3mm}+\frac{k_{j}\varphi(k)}{\omega_{m}(k)\sqrt{2\omega(k)}}\big(E_{0}(H_{m})-H_{m}-\omega_{m}(k)\big)^{-2}(\mu L_{1}(k)+2\lambda\overline{L_{2}(k)})\Phi_{m}
\\
&\hspace{3mm}-\frac{i\varphi(k)}{\sqrt{2\omega(k)}}\big(E_{0}(H_{m})-H_{m}-\omega_{m}(k)\big)^{-1}\big(\mu L_{1.j}(k)+2\lambda\overline{L_{2.j}(k)}\big)\Phi_{m},
\end{aligned}
\end{equation*}
where
\begin{equation*}
\begin{aligned}
S_{1.j}&:=\displaystyle\int_{\mathbb{R}^{d}}x_{j}\chi_{\textrm{sp}}(x)e^{-ikx}\phi(f_{x}) \textrm{d}x\hspace{5mm}S_{2,j}:=\displaystyle\int_{\mathbb{R}^{d}}x_{j}\chi_{\textrm{sp}}(x)\phi(f_{x})\phi(f_{x})^{\ast}\phi(f_{x}) \textrm{d}x
\\
L_{1.j}&:=\displaystyle\int_{\mathbb{R}^{d}}x_{j}\chi_{\textrm{sp}}(x)e^{-ikx}\phi(f_{x})^{\ast} \textrm{d}x\hspace{5mm}L_{2,j}:=\displaystyle\int_{\mathbb{R}^{d}}x_{j}\chi_{\textrm{sp}}(x)\phi(f_{x})^{\ast}\phi(f_{x})\phi(f_{x})^{\ast} \textrm{d}x,
\end{aligned}
\end{equation*}
and $D_{j}$ is the strong differential operator in the $j$-th variable $k_{j}$.
\vspace{2mm}
\\
\textit{Proof.}
Since $(E_{0}(H_{m})-H_{m}-\omega_{m}(\cdot))$ is differentiable in operator norm topology and $\varphi/\sqrt{\omega}$ is differentiable for any $k\neq0$, it suffices to show the strong differentiability of $S_{1}$, $S_{2}$, $L_{1}$ and $L_{2}$. Here we only show the case of $S_{2}$. Since $\Phi_{m}\in D(H_{m})$, we can take a sequence $\{\Phi_{m}^{j}\}_{j=1}^{\infty}\subset \mathscr{F}_{\textrm{b,fin}}([C_{0}^{\infty}(\mathbb{R}^{d})])$ such that $\Phi_{m}^{j}\rightarrow \Phi_{m}$ and $H_{m}\Phi_{m}^{j}\rightarrow H_{m}\Phi_{m}\hspace{2mm}(j\rightarrow\infty)$. Since $S_{2}(k)$ and $S_{2,l}(k)$ is $H_{m}$-bounded, we have $S_{2}(k)\Phi_{m}^{j}\rightarrow S_{2}(k)\Phi_{m}$ and $S_{2,l}(k)\Phi_{m}^{j}\rightarrow \overline{S_{2,l}(k)}\Phi_{m}$\hspace{2mm}$(j\rightarrow\infty)$. Let $\{e_{l}\}_{l=1}^{d}$ be the standard orthogonal basis of $\mathbb{R}^{d}$ and $h\in \mathbb{R}\setminus\{0\}$. It is seen that
\begin{equation*}
\begin{aligned}
&\hspace{5mm}\big\|\frac{\overline{S_{2}(k+he_{l})}-\overline{S_{2}(k)}}{h}\Phi_{m}+i\overline{S_{2,l}(k)}\Phi_{m}\big\|^{2}
\\
&=\displaystyle\lim_{j\rightarrow\infty}\big\|\frac{S_{2}(k+he_{l})-S_{2}(k)}{h}\Phi^{j}_{m}+iS_{2,l}(k)\Phi^{j}_{m}\big\|^{2}
\\
&\le\displaystyle\lim_{j\rightarrow\infty}\displaystyle\int_{\mathbb{R}^{d}\times\mathbb{R}^{d}}\chi_{\textrm{sp}}(x)\chi_{\textrm{sp}}(y)\big|\frac{e^{ihx_{l}}-1}{h}-ix_{l}\big|\big|\frac{e^{-ihy_{l}}-1}{h}+iy_{l}\big|\big|\langle \phi(f_{x})\phi(f_{x})^{\ast}\phi(f_{x})\Phi_{m}^{j}, \phi(f_{y})\phi(f_{y})^{\ast}\phi(f_{y})\Phi_{m}^{j}\rangle\big| \textrm{d}x\textrm{d}y
\end{aligned}
\end{equation*}
\begin{align}
&\le\displaystyle\lim_{j\rightarrow\infty}\displaystyle\int_{\mathbb{R}^{d}}\chi_{\textrm{sp}}(x)\big|\frac{e^{ihx_{l}}-1}{h}-ix_{l}\big|^{2}\textrm{d}x
\notag \\
&\hspace{30mm}\times\Big(\displaystyle\int_{\mathbb{R}^{d}\times\mathbb{R}^{d}}\chi_{\textrm{sp}}(x)\chi_{\textrm{sp}}(y)\big|\langle\phi(f_{x})\phi(f_{y})^{\ast}\Phi_{m}^{j}, \phi(f_{x})^{\ast}\phi(f_{x})\phi(f_{y})^{\ast}\phi(f_{y})\Phi_{m}^{j}\rangle\big|^{2} \textrm{d}x\textrm{d}y\Big)^{1/2}
\\
&\le \displaystyle\lim_{j\rightarrow\infty}C\|(d\Gamma_{\textrm{b}}([\omega_{m}])+1)\Phi_{m}^{j}\|\displaystyle\int_{\mathbb{R}^{d}}\chi_{\textrm{sp}}(x)\big|\frac{e^{ihx_{l}}-1}{h}-ix_{l}\big|^{2} \textrm{d}x\textrm{d}y
\notag \\
&\hspace{30mm}\times\Big(\displaystyle\int_{\mathbb{R}^{d}\times\mathbb{R}^{d}}\chi_{\textrm{sp}}(x)\chi_{\textrm{sp}}(y)\langle\big(\phi(f_{x})^{\ast}\phi(f_{x})\big)^{2}\Phi_{m}^{j}, \big(\phi(f_{y})^{\ast}\phi(f_{y})\big)^{2}\Phi_{m}^{j}\rangle \textrm{d}x\textrm{d}y\Big)^{1/2}
\notag \\
&\le \displaystyle\lim_{j\rightarrow\infty}C\big\|(\textrm{d}\Gamma_{\textrm{b}}([\omega_{m}])+1)\Phi_{m}^{j}\big\|\big\|H_{2}\Phi_{m}^{j}\big\|\displaystyle\int_{\mathbb{R}^{d}}\chi_{\textrm{sp}}(x)\big|\frac{e^{ihx_{l}}-1}{h}-ix_{l}\big|^{2} \textrm{d}x.
\end{align}
Here to get (26),  we used the Schwartz inequality. Since $\textrm{d}\Gamma_{\textrm{b}}([\omega_{m}])$ and $H_{2}$ are $H_{m}$-bounded, the limit of (27) exists and is independent of $h$. Note that 
$|(e^{ihx_{l}}-1)/h-ix_{l}|^{2}\le 4x_{l}^{2}$
and $\int_{\mathbb{R}^{d}}\chi_{\textrm{sp}}(x)x_{l}^{2}<\infty$. Hence from Lebesgue's dominated convergence theorem, we see that $\overline{S_{2}(k)}\Phi_{m}$ is strongly differentiable and its strong derivative is $-i\overline{S_{2,l}}(k)\Phi_{m}$. By using the Leibniz rule for (\ref{ptf}), we obtain the desired results. \hspace{30mm}$\square$
\vspace{2mm}
\\
\textbf{Lemma 5.7.}
\textit{Suppose that the same assumption as in Lemma 5.6 holds. Then there exist constants $C_{1}, C_{2}$ and $C_{3}>0$ independent of $m$ such that}
\begin{equation}\label{bound1}
\big\|D_{j}a_{\pm}(k)\Phi_{m}\big\|_{\mathscr{H}}\le C_{1}\frac{|\varphi(k)|}{\omega(k)^{3/2}}+C_{2}\frac{|\varphi(k)|}{\omega(k)^{5/2}}+C_{3}\frac{|(\partial_{j}\varphi)(k)|}{\omega(k)^{3/2}}\hspace{5mm}\text{for}\hspace{2mm}k\neq0.
\end{equation}
 \textit{Moreover, under the additional assumption that} $\varphi\in D(\omega^{-5/2})$ \textit{and} $\nabla_{k}\varphi\in D(\omega^{-3/2})$, \textit{one has}
\begin{equation}\label{bound2}
\displaystyle\sup_{0<m\le1}\displaystyle\sum_{j=1}^{d}\displaystyle\int_{\mathbb{R}^{d}}\big\|D_{j} a_{\pm}(k)\Phi_{m}\big\|^{2}_{\mathscr{H}}\textrm{d}k<\infty.
\end{equation}
\begin{proof}
For $k\neq0$, it is seen that $\big\|(E_{0}(H_{m})-H_{m}-\omega_{m}(k))^{-1}\big\|\le \omega(k)^{-1}$. Since $S_{1}(k)$, $S_{2}(k)$, $L_{1}(k)$, $L_{2}(k)$, $S_{1.j}(k)$, $S_{2.j}(k)$, $L_{1,j}(k)$ and $L_{2,j}$ are $H_{m}$-bounded and its bound are independent of $k$, we have (\ref{bound1}). (\ref{bound2}) is immediately follows from (\ref{bound1}).
\end{proof}
We set $\Phi_{m}=\{\Phi_{m}^{(n,n')}\}_{n,n'=0}^{\infty}$. Note that $\Phi_{m}^{(n,n')}$ is $d(n+n')$-variable function. We denote $k_{j}=(k_{j.1},\dots k_{j.d})$ and $l_{j}=(l_{j.1},\dots ,l_{j.d})$.
\vspace{2mm}
\\
\textbf{Lemma 5.8.} \textit{For} $1\le i\le n$ \textit{and} $1\le j\le d$, \textit{let} $\partial_{i,j}$ \textit{be the distributional derivative in} $k_{i,j}$ \textit{in} $\Omega$ \textit{and for} $n+1\le i\le n+n'$ \textit{and} $1\le j\le d$, $\partial_{i.j}$ \textit{be the distributional derivative in} $l_{i.j}$ \textit{in} $\Omega$. \textit{Suppose that Assumption 2.2 holds. Then},
\begin{equation*}
(\partial_{i.j}\Phi_{m}^{(n,n')})(k_{1},\dots,k_{n}:l_{1},\dots,l_{n'})= \begin{cases}
                                                                               \frac{1}{\sqrt{n}}D_{j}a_{+}(k_{i})\Phi_{m}^{(n-1,n')}(k_{1},\dots, \hat{k_{i}},\dots, k_{n}:l_{1},\dots ,l_{n'}) & 1\le i\le n ,
\\
\frac{1}{\sqrt{n'}}D_{j}a_{-}(l_{i-n})\Phi_{m}^{(n,n'-1)}(k_{1},\dots, k_{n}:l_{1},\dots, \hat{l_{i}},\dots, l_{n'})& n+1\le i\le n+n',

                                                                           \end{cases}
\end{equation*}
where $\hat{k}$ denotes omitting of $k$. 
\begin{proof}
Here, we consider only the case of $1\le i\le n$ and $1\le j\le d$. The other case is proven in a similar manner. Let $f\in C_{0}^{\infty}(\Omega^{n+n'})$. Then it suffices to show that
\begin{equation}\label{goal}
\begin{aligned}
\displaystyle\int_{\mathbb{R}^{d(n+n')}}&\Phi_{m}^{(n,n')}(k_{1},\dots ,k_{n}:l_{1},\dots ,l_{n})(\partial_{i.j}f)(k_{1},\dots ,k_{n},l_{1},\dots, l_{n'})\textrm{d}^{n}k\textrm{d}^{n'}l
\\
&+\frac{1}{\sqrt{n}}\displaystyle\int_{\mathbb{R}^{d(n+n')}}D_{j}a_{+}(k_{i})\Phi_{m}^{(n-1,n')}(k_{1},\dots, \hat{k_{i}},\dots, k_{n}:l_{1},\dots ,l_{n'})f(k_{1},\dots k_{n},l_{1},\dots, l_{n'}) \textrm{d}^{n}k\textrm{d}^{n'}l=0,
\end{aligned}
\end{equation}
where $\textrm{d}^{n}k:=\textrm{d}k_{1}\cdots\textrm{d}k_{n}$, $\textrm{d}^{n'}l:=\textrm{d}l_{1}\cdots\textrm{d}l_{n'}$. We denote the standard orthogonal basis of $\mathbb{R}^{d}$ by $\{e_{j}\}_{j=1}^{d}$. By the definition of classical derivative, the left hand side of (\ref{goal}) is calculated as follows:
\begin{equation*}
\begin{aligned}
\displaystyle\lim_{h\rightarrow0}\Big|\displaystyle\int_{\mathbb{R}^{d(n+n')}}&\frac{\Phi_{m}^{(n,n')}(k_{1},\dots, k_{i}+he_{j},\dots k_{n}: l_{1},\dots, l_{n'})-\Phi_{m}^{(n,n')}(k_{1},\dots, k_{n}:l_{1},\dots, l_{n'})}{h}f(k_{1},\dots, k_{n},l_{1},\dots l_{n'})\textrm{d}^{n}k\textrm{d}^{n'}l
\\
&-\frac{1}{\sqrt{n}}\displaystyle\int_{\mathbb{R}^{d(n+n')}}D_{j}a_{+}(k_{i})\Phi_{m}^{(n-1,n')}(k_{1},\dots, \hat{k_{i}},\dots, k_{n}:l_{1},\dots ,l_{n'})f(k_{1},\dots k_{n},l_{1},\dots, l_{n'}) \textrm{d}^{n}k\textrm{d}^{n'}l\Big|.
\\
\end{aligned}
\end{equation*}
Since $\Phi_{m}^{(n,n')}\in L^{2}_{\textrm{sym}}(\mathbb{R}^{dn}\times\mathbb{R}^{dn'})$, we have
\begin{equation*}
\begin{aligned}
\displaystyle\lim_{h\rightarrow0}\frac{1}{\sqrt{n}}\Big|\displaystyle\int_{\mathbb{R}^{d(n+n')}}&\frac{\big(a_{+}(k_{i}+he_{j})-a_{+}(k_{i})\big)\Phi_{m}^{(n-1,n')}(k_{1},\dots, k_{i-1}, k_{i+1},\dots k_{n}: l_{1},\dots, l_{n'})}{h}f(k_{1},\dots, k_{n},l_{1},\dots l_{n'})\textrm{d}^{n}k\textrm{d}^{n'}l
\\
&-\displaystyle\int_{\mathbb{R}^{d(n+n')}}D_{j}a_{+}(k_{i})\Phi_{m}^{(n-1,n')}(k_{1},\dots, \hat{k_{i}},\dots, k_{n}:l_{1},\dots ,l_{n'})f(k_{1},\dots k_{n},l_{1},\dots, l_{n'}) \textrm{d}^{n}k\textrm{d}^{n'}l\Big|.
\end{aligned}
\end{equation*}
By applying the Schwarz inequality with respect to $\textrm{d}k_{1}\cdots\textrm{d}k_{i-1}\textrm{d}k_{i+1}\cdots\textrm{d}k_{n}\textrm{d}^{n'}l$, we see that it is dominated by
\begin{equation}\label{aaaa}
\displaystyle\lim_{h\rightarrow0}\frac{1}{\sqrt{n}}\displaystyle\int_{\mathbb{R}^{d}}\big\|\frac{\big(a_{+}(k_{i}+he_{j})-a_{+}(k_{i})\big)\Phi_{m}^{(n-1,n')}}{h}-D_{j}a_{+}(k_{i})\Phi_{m}^{(n-1,n')}\big\|_{L^{2}(\mathbb{R}^{d(n+n'-1)})}\big\|f(\cdot,k_{i},\cdot)\big\|_{L^{2}(\mathbb{R}^{d(n+n'-1)})}\textrm{d}k_{i}.
\end{equation}
Since the function $k\mapsto a_{+}(k)\Phi_{m}^{(n-1,n')}$ is strongly continuous differentiable in $\Omega$, the first factor of the integrand of  (\ref{aaaa}) is bounded on $\Omega$ uniformly in $h$. Therefore, we can apply the Lebesgue dominated convergence theorem and the desired result follows. 
\end{proof}
Let us denote the Sobolev space of order 1 and index $p$ on open set $U$ in $\mathbb{R}^{d(n+n')}$ by $W^{1,p}(U)$. 
\vspace{2mm}
\\
\textbf{Lemma 5.9.}
\textit{Suppose that Assumption 2.2 holds. Then for any $n+n'\ge1$, $0<m\le1$ and $1\le p <2$ , $\Phi_{m}^{(n,n')}\in W^{1,p}(\Omega^{n+n'})$ and}
\begin{equation*}
\displaystyle\sup_{0<m\le1}\big\|\Phi_{m}^{(n,n')}\big\|_{W^{1,p}(\Omega^{n+n'})}<\infty.
\end{equation*}
\begin{proof}
Similar to the proof of [14, Proof of Theorem 2.1, Step 2]. To prove this, we need Lemma 5.7 and Lemma 5.8.
\end{proof}
\begin{proof}[Proof of Theorem 2.3] 
Since $\{\Phi_{m}\}_{0<m\le1}$ is a bounded set on $\mathscr{H}$, there exists a sequence $\{\Phi_{m_{j}}\}_{j=1}^{\infty}$ and a vector $\Phi\in \mathscr{H}$ such that $m_{j}\rightarrow0$ as $j\rightarrow\infty$ and 
\begin{equation*}
\displaystyle\wlim_{j\rightarrow\infty}\Phi_{m_{j}}=\Phi.
\end{equation*}
Let $z\in \mathbb{C}\setminus \mathbb{R}$ and $\Psi\in \mathscr{H}$ be arbitrary. Then
\begin{equation}\label{gs}
\langle \Psi, (H_{m_{j}}-z)^{-1}\Phi_{m_{j}}\rangle=\langle \Psi, (E_{0}(H_{m_{j}})-z)^{-1}\Phi_{m_{j}}\rangle.
\end{equation} 
By taking the limit $j\rightarrow \infty$ on the both side of (\ref{gs}), we have by Lemma 5.3, 
\begin{equation*}
\langle\Psi,(H-z)^{-1}\Phi\rangle=\langle \Psi, (E_{0}(H)-z)^{-1}\Phi\rangle.
\end{equation*}
This fact implies that $\Phi\in D(H)$ and
\begin{equation*}
H\Phi=E_{0}(H)\Phi.
\end{equation*}
Hence $\Phi$ is a ground state of $H$ if $\Phi\neq0$. Now we assume that $\Phi=0$. By Lemma 5.5, we have
\begin{equation}\label{ggs}
\big\|\Phi_{m_{j}}\big\|^{2}=\displaystyle\sum_{n+n'\le N}\big\|\Phi_{m_{j}}^{(n,n')}\big\|^{2}+\displaystyle\sum_{n+n'> N}\big\|\Phi^{(n,n')}_{m_{j}}\big\|^{2}\le\displaystyle\sum_{n+n'\le N}\big\|\Phi^{(n,n')}_{m_{j}}\big\|^{2}+\frac{1}{N}\displaystyle\sup_{0<m\le1}\big\|N_{\textrm{b}}^{1/2}\Phi_{m}\big\|^{2},
\end{equation}
where $N\in\mathbb{N}$ is arbitrary. Here we show that for any $n,n'$, $\Phi_{m_{j}}^{(n,n')}$ converges to $\Phi^{(n,n')}$ strongly. By applying Lemma 5.4 and the definition of annihilation kernel, we have 
\begin{equation*}
\text{supp}\hspace{1mm}\Phi_{m_{j}}^{(n,n')}=\overline{\Omega^{n+n'}},
\end{equation*}
since $\Phi_{m_{j}}^{(n,n')}\in L^{2}_{\textrm{sym}}(\mathbb{R}^{dn}\times\mathbb{R}^{dn'})$(see ,e.g., [14, Proof of Theorem 2.1, Step2]). Here, $\overline{A}$ denotes the closure of $A\subset \mathbb{R}^{d(n+n')}$. Since the Lebesgue measure of $\Omega^{n+n'}$ is finite, we have $L^{s}(\Omega^{n+n'})\subset L^{2}(\Omega^{n+n'})$ for all $s\ge2$. Therefore, $\Phi_{m_{j}}^{(n,n')}$ weakly converges to $\Phi^{(n,n')}=0$ in the $L^{p}(\Omega^{n+n'})$ sense. By Lemma 5.9, a subsequence of $\{\Phi_{m_{j}}\}_{j=1}^{\infty}$ converges to a vector $\hat{\Phi}^{(n,n')}\in W^{1,p}(\Omega^{n+n'})$ in the $W^{1,p}(\Omega^{n+n'})^{\ast}$ sense. It means that for any $f_{0}, f_{1}, \dots, f_{d(n+n')}\in L^{p}(\Omega^{n+n'})^{\ast}=L^{s}(\Omega^{n+n'})$ with $1/s+1/p=1$, 
\begin{equation*}
\displaystyle\int_{\Omega^{n+n'}}f_{0}\big(\Phi^{(n,n')}_{m_{j}}-\hat{\Phi}^{(n,n')}\big) \textrm{d}^{n}k\textrm{d}^{n'}l+\displaystyle\sum_{i=1}^{d(n+n')}\int_{\Omega^{n+n'}}f_{i}\partial_{i}\big(\Phi^{(n,n')}_{m_{j}}-\hat{\Phi}^{(n,n')}\big) \textrm{d}^{n}k\textrm{d}^{n'}l\rightarrow 0,
\hspace{5mm}j\rightarrow\infty.
\end{equation*}
Hence we have
\begin{equation*}
0=\Phi^{(n,n')}(k_{1},\dots ,k_{n}:l_{1}\dots ,l_{n'})=\hat{\Phi}^{(n,n')}(k_{1},\dots, k_{n}:l_{1},\dots, l_{n'})\hspace{3mm}a.e.
\end{equation*}
Thus we have for all $1\le p<2$, $\Phi_{m_{j}}^{n,n'}\rightarrow 0$ as $j\rightarrow\infty$ in the weak sense of $W^{1,p}(\Omega^{n+n'})$. By applying the Rellich-Kondrachov theorem (see, e.g.,[1, Theorem 6.3],[18, Theorem 8.9]),  we have
\begin{equation*}
\displaystyle\lim_{j\rightarrow\infty}\big\|\Phi_{m_{j}}^{(n,n')}\big\|_{L^{q}(\Omega^{n+n'})}=0,
\end{equation*}
for all $q<\frac{d(n+n')p}{d(n+n')-p}$, since $\Omega$ has cone property. To get $q=2$, we choose $p$ as 
\begin{equation*}
\begin{cases}
&\frac{2d(n+n')}{d(n+n')+2}<p<2,\hspace{2mm}\text{if}\hspace{2mm}2\le d(n+n'),
\\
& \hspace{8mm}p=1,\hspace{13mm}\text{if}\hspace{2mm}d(n+n')=1.
\end{cases}
\end{equation*}
Thus, by taking $\limsup_{j\rightarrow\infty}$ in (\ref{ggs}), we have
\begin{equation*}
1=\limsup_{j\rightarrow\infty}\big\|\Phi_{m_{j}}\big\|^{2} \le \frac{1}{N}\displaystyle\sup_{0<m\le1}\big\|N_{\textrm{b}}^{1/2}\Phi_{m}\big\|^{2}.
\end{equation*}
But this is a contradiction since $N$ is arbitrary. Hence $\Phi\neq0$.
\end{proof}
\section{Total charge of a ground state}
In this section, we prove Theorem 2.4 and 2.5.  
\vspace{2mm}
\\
\textit{Proof of Theorem 2.4}
\hspace{3mm}It is trivial that $H_{0}$ and $e^{-itQ}$ commute (see Proposition A.4). By Proposition A.2-(2) and Proposition A.4-(2),  following relations hold:
\begin{equation*}
e^{-itQ}a_{+}(u)e^{itQ}=a_{+}(e^{-itq}u),\hspace{5mm}e^{-itQ}a_{-}(u)e^{itQ}=a_{-}(e^{itq}u),
\end{equation*}
\begin{equation*}
e^{-itQ}a_{+}(u)^{\ast}e^{itQ}=a_{+}(e^{-itq}u)^{\ast},\hspace{5mm}e^{-itQ}a_{-}(u)^{\ast}e^{itQ}=a_{-}(e^{itq}u)^{\ast},\hspace{3mm}u\in L^{2}(\mathbb{R}^{d}).
\end{equation*}
 Let $\Psi\in\mathscr{F}_{\textrm{b,fin}}([C_{0}^{\infty}(\mathbb{R}^{d})])$. Then, $e^{itQ}\Psi\in \mathscr{F}_{\textrm{b,fin}}([C_{0}^{\infty}(\mathbb{R}^{d})])$ and we have
\begin{equation*}
\begin{aligned}
e^{-itQ}H_{1}e^{itQ}\Psi&=\displaystyle\int_{\mathbb{R}^{d}}\chi_{\textrm{sp}}(x)e^{-itQ}(\phi(f_{x})^{\ast}\phi(f_{x}))e^{itQ}\Psi dx,
\\
e^{-itQ}H_{2}e^{itQ}\Psi&=\displaystyle\int_{\mathbb{R}^{d}}\chi_{\textrm{sp}}(x)e^{-itQ}(\phi(f_{x})^{\ast}\phi(f_{x}))^{2}e^{itQ}\Psi dx.
\end{aligned}
\end{equation*}
It follows that on $\mathscr{F}_{\textrm{b,fin}}([C_{0}^{\infty}(\mathbb{R}^{d})])$:
\begin{equation*}
\begin{aligned}
e^{-itQ}\phi(f_{x})^{\ast}\phi(f_{x})e^{itQ}&=\frac{1}{2}(a_{+}(e^{-itq}f_{x})^{\dag}+a_{-}(e^{itq}f_{x}))(a_{+}(e^{-itq}f_{x})+a_{-}(e^{itq}f_{x})^{\dag})
\\
&=e^{-itq}\phi(f_{x})^{\ast}e^{itq}\phi(f_{x})=\phi(f_{x})^{\ast}\phi(f_{x}).
\end{aligned}
\end{equation*}
Therefore for any $\Psi\in\mathscr{F}_{\textrm{b,fin}}([C_{0}^{\infty}(\mathbb{R}^{d})])$, we see that
\begin{equation*}
e^{-itQ}He^{itQ}\Psi=H\Psi.
\end{equation*}
Since $e^{-itQ}$ is unitary and $\mathscr{F}_{\textrm{b,fin}}([C_{0}^{\infty}(\mathbb{R}^{d})])$ is a core of $H$, above equality can be extended to operator equality. By the functional calculus, we have
\begin{equation*}
e^{-itQ}e^{-isH}e^{itQ}=e^{-isH},\hspace{5mm}(s,t\in\mathbb{R}).
\end{equation*}
Hence the desired result follows.\hspace{120mm}$\square$
\vspace{2mm}
\\
\textbf{Remark 6.1.} Also the \textit{massive} Hamiltonian $H_{m}$ strongly commutes with $Q$. The proof is quite similar to that of Theorem 2.4.
\vspace{2mm}
\\
\textbf{Lemma 6.1.} \textit{We assume Assumption 2.1 and 2.2. Then for} $k\neq0$, \textit{it follows that}
\begin{equation*}
\begin{aligned}
&a_{+}(k)\Phi_{g}=\frac{\varphi(k)}{\sqrt{2\omega(k)}}(E_{0}(H)-H-\omega(k))^{-1}(\mu S_{1}(k)+2\lambda \overline{S_{2}(k)})\Phi_{g},
\\
&a_{-}(k)\Phi_{g}=\frac{\varphi(k)}{\sqrt{2\omega(k)}}(E_{0}(H)-H-\omega(k))^{-1}(\mu L_{1}(k)+2\lambda \overline{L_{2}(k)})\Phi_{g}.
\end{aligned}
\end{equation*}
$Especially$, $\Phi_{g}\in D(N_{\textrm{b}}^{1/2})$.
\vspace{2mm}
\\
\hspace{5mm}Since the proof of this lemma is quite similar to that of Lemma 5.4 and Lemma 5.5,  we omit it.
\vspace{2mm}
\\
\textit{Proof of Theorem 2.5.}\hspace{2mm}Let $N_{+}:=d\Gamma_{\textrm{b}}(1)\otimes1$, $N_{-}:=1\otimes d\Gamma_{\textrm{b}}(1)$ and  $U$ be the canonical unitary operator acting from $\mathscr{H}$ to $\mathscr{F}_{\textrm{b}}(L^{2}(\mathbb{R}^{d}))\otimes\mathscr{F}_{\textrm{b}}(L^{2}(\mathbb{R}^{d}))$ (see Appendix A). By Proposition A.3, we have
\begin{equation*}
UN_{\textrm{b}}U^{-1}=N_{+}+N_{-},\hspace{3mm}UQU^{-1}=q(\overline{N_{+}-N_{-}}).
\end{equation*}
Suppose that $\Phi_{g}\in \mathscr{H}_{q}(z)$ for some $z$ with $|z|\ge n_{0}$. 
Note that $Q\Phi_{g}=zq\Phi_{g}$ and $\|\Phi_{g}\|=1$. Then
\begin{equation*}
|zq|=|\langle \Phi_{g}, Q\Phi_{g}\rangle|=|q||\langle U\Phi_{g},(\overline{N_{+}-N_{-}})U\Phi_{g}\rangle|\le|q|\big\|N_{+}^{1/2}U\Phi_{g}\big\|^{2}+|q|\big\|N_{-}^{1/2}U\Phi_{g}\big\|^{2}=|q|\big\|N_{\textrm{b}}^{1/2}\Phi_{g}\big\|^{2}<n_{0}|q|.
\end{equation*}
Thus we have $|z|<n_{0}$. But this is a contradiction. Hence $\Phi_{g}\notin\mathscr{H}_{q}(z)$ for all $|z|\ge n_{0}$.\hspace{39mm}$\square$
\vspace{2mm}
\\
\textbf{Concluding remark}
One of the next tasks is to analyze the Hamiltonian $H$ on each fixed total charge space $\mathscr{H}_{q}(z)$ with $z\in\mathbb{Z}$.  We leave it for future study.
\vspace{5mm}
\\
\textbf{\Large{APPENDIX A}}
\\
\hspace{3mm}In this section, we introduce some facts which are often used in this paper and are well known. We use the same notations and symbols as in Section 2. Let $\mathcal{X}$ and $\mathcal{Y}$ be Hilbert spaces.
\vspace{3mm}
\\
\textbf{Proposition A.1.}[\textit{3, Proposition 4.24}][\textit{4, Lemma 6.32}]
 \textit{Let $T$ be a non-negative self-adjoint operator on $\mathcal{X}$ with ker $T=\{0\}$. If $u\in D(T^{-1/2})$, then}
\begin{equation*}
\begin{aligned}
\big\|A(u)\Psi\big\|&\le\big\|T^{-1/2}u\big\|\big\|\textrm{d}\Gamma_{\textrm{b}}(T)^{1/2}\Psi\big\|,
\\
\big\|A(u)^{\dag}\Psi\big\|&\le\big\|T^{-1/2}u\big\|\big\|\textrm{d}\Gamma_{\textrm{b}}(T)^{1/2}\Psi\big\|+\big\|u\big\|\big\|\Psi\big\|,
\end{aligned}
\end{equation*}
\textit{for all} $\Psi\in D(\textrm{d}\Gamma_{\textrm{b}}(T)^{1/2})$. \textit{Moreover if} $u,v\in D(T)\cap D(T^{-1/2})$, then
\begin{equation*}
\big\|A(u)^{\sharp}A(v)^{\natural}\Psi\big\| \le C \big\|(\textrm{d}\Gamma_{\textrm{b}}(T)+1)\Psi\big\|\big(\big\|T^{-1/2}u)\big\|+\big\|u\big\|\big)\big(\big\|T^{-1/2}v\big\|+\big\|v\big\|+\big\|Tv\big\|+\big\|T^{1/2}v\big\|\big),
\end{equation*}
\textit{for all} $\Psi\in D(\textrm{d}\Gamma_{\textrm{b}}(T))$. \textit{Here} $C>0$ \textit{is a constant independent of $u, v,T$ and $\Psi$}.
\vspace{3mm}
\\
\textbf{Proposition A.2.}[\textit{3, Proposition 4.26}][\textit{8, Lemma 2.7 and Lemma 2.8}]
\textit{Let $T$ be a densely defined closable operator on $\mathcal{X}$, and $u\in D(T)\cap D(T^{\ast})$}. Then:
\vspace{2mm}
\\
\hspace{3mm}(1)
\begin{equation*}
[\textrm{d}\Gamma_{\textrm{b}}(T), A(u)] = -A(T^{\ast}u),\hspace{7mm}\text{and}\hspace{7mm}[\textrm{d}\Gamma_{\textrm{b}}(T), A(u)^{\dag}] =A(Tu)^{\dag},\hspace{3mm}\text{on $\mathscr{F}_{\textrm{b,fin}}(D(T))$}.
\end{equation*}
\hspace{3mm}(2)
 \textit{If $u\in D(T)$, then}
\begin{equation*}
\Gamma_{\textrm{b}}(T)A(u)^{\dag}=A(Tu)^{\dag}\Gamma_{\textrm{b}}(T),\hspace{5mm}\text{on}\hspace{2mm}\mathscr{F}_{\text{b,fin}}(D(T)).
\end{equation*}
\hspace{5mm}\textit{Moreover, if $T$ is isometry, then}
\begin{equation*}
\Gamma_{\textrm{b}}(T)A(u)=A(Tu)\Gamma_{\textrm{b}}(T).
\end{equation*}
\textbf{Proposition A.3.}[\textit{3, Theorem 4-55 and Theorem 4-56}]
\textit{Let $\mathcal{X}$ and $\mathcal{Y}$ be Hilbert spaces. Then there exists a unique unitary operator} $U_{\mathcal{X},\mathcal{Y}}$: $\mathscr{F}_{\textrm{b}}(\mathcal{X}\oplus\mathcal{Y})\rightarrow \mathscr{F}_{\textrm{b}}(\mathcal{X})\otimes\mathscr{F}_{\textrm{b}}(\mathcal{Y})$ \textit{such that the following (1) and (2) are hold:}
\vspace{2mm}
\\
\hspace{3mm}(1)
\begin{equation*}
U_{\mathcal{X},\mathcal{Y}}\Omega_{\mathcal{X}\oplus\mathcal{Y}}=\Omega_{\mathcal{X}}\otimes\Omega_{\mathcal{Y}},
\end{equation*}
\hspace{5mm}\textit{where} $\Omega_{\mathcal{X}}$ \textit{is the Fock vacuum in} $\mathscr{F}_{\textrm{b}}(\mathcal{X})$.
\vspace{2mm}
\\
\hspace{2mm}(2)
\begin{equation*}
U_{\mathcal{X},\mathcal{Y}}A(u\oplus v)^{\sharp}U_{\mathcal{X},\mathcal{Y}}^{-1}=\overline{A(u)^{\sharp}\otimes I+I\otimes A(v)^{\sharp}},\hspace{4mm}u\in\mathcal{X}, v\in\mathcal{Y},
\end{equation*}
\hspace{5mm}\textit{and}
\begin{equation*}
U_{\mathcal{X},\mathcal{Y}}\mathscr{F}_{\textrm{b,fin}}(\mathcal{X}\oplus\mathcal{Y})=\mathscr{F}_{\textrm{b,fin}}(\mathcal{X})\hat{\otimes}\mathscr{F}_{\textrm{b,fin}}(\mathcal{Y}),
\end{equation*}
\hspace{5mm}\textit{where $A(\cdot)^{\sharp}$ denotes $A(\cdot)$ or $A(\cdot)^{\dag}$}. \textit{Moreover, for all self-adjoint operators $T$ on $\mathcal{X}$ and $S$ on} $\mathcal{Y}$,
\begin{equation*}
U_{\mathcal{X},\mathcal{Y}}\textrm{d}\Gamma_{\textrm{b}}(T\oplus S)U^{-1}_{\mathcal{X},\mathcal{Y}}=\overline{\textrm{d}\Gamma_{\textrm{b}}(T)\otimes I+I\otimes \textrm{d}\Gamma_{\textrm{b}}(S)}.
\end{equation*}
\vspace{3mm}
\textbf{Remark} \hspace{2mm}\textit{If $T$ and $S$ are non-negative in the above}, 
\begin{equation*}
\overline{\textrm{d}\Gamma_{\textrm{b}}(T)\otimes I+I\otimes \textrm{d}\Gamma_{\textrm{b}}(S)}=\textrm{d}\Gamma_{\textrm{b}}(T)\otimes I+I\otimes \textrm{d}\Gamma_{\textrm{b}}(S).
\end{equation*}
\vspace{3mm}
\textbf{Proposition A.4.}[\textit{3, Theorem 4-17 and Theorem 4-20}]
\textit{Let $A$ and $B$ are self-adjoint on $\mathscr{K}$}.
\\
\hspace{3mm}(1) \textit{$A$ and $B$ are strongly commute if and only if} $\textrm{d}\Gamma_{\textrm{b}}(A)$ $and$ $\textrm{d}\Gamma_{\textrm{b}}(B)$ \textit{are strongly commute}.
\vspace{2mm}
\\
\hspace{3mm}(2)
\begin{equation*}
\Gamma_{\textrm{b}}(e^{-itA})=e^{-it\textrm{d}\Gamma_{\textrm{b}}(A)}.
\end{equation*}
Let $\mathcal{K}=L^{2}(\mathbb{R}^{d})$. Then $\mathscr{F}_{\textrm{b}}(L^{2}(\mathbb{R}^{d}))$ is written as follows:
\begin{equation*}
\mathscr{F}_{\textrm{b}}(L^{2}(\mathbb{R}^{d}))=\mathbb{C}\oplus\displaystyle\bigoplus_{n=1}^{\infty}L^{2}_{\textrm{sym}}(\mathbb{R}^{dn}),
\end{equation*}
where
\begin{equation*}
L^{2}_{\textrm{sym}}(\mathbb{R}^{dn}):=\big\{f\in L^{2}(\mathbb{R}^{dn}): f(k_{\pi(1)},\cdots,k_{\pi(n)})=f(k_{1},\cdots,k_{n}) \hspace{2mm}\text{for a,e, $k_{1}, \cdots , k_{n} \in\mathbb{R}^{d}$ and $\pi\in \mathcal{S}_{n}$}\big\}.
\end{equation*}
For a,e, $k\in\mathbb{R}^{d}$, an annihilation kernel $a(k)$ act on $\mathscr{F}_{\textrm{b}}(L^{2}(\mathbb{R}^{d}))$ is defined as follows.
\begin{equation*}
(a(k)\Psi)^{(n)}(k_{1},\cdots, k_{n}):=\sqrt{n+1} \Psi^{(n+1)}(k, k_{1},\cdots, k_{n}).
\end{equation*}
\textbf{Proposition A.5.}[\textit{3, Proposition 8.6}]\hspace{2mm}\textit{Let $f$ be a measurable function such that $0\le f(k)<\infty$ for a,e,}$k\in\mathbb{R}^{d}$.
\textit{Then} $\Psi\in D(\textrm{d}\Gamma_{\textrm{b}}(f)^{1/2})$ \textit{if and only if} 
\begin{equation*}
\displaystyle\int_{\mathbb{R}^{d}}f(k)\big\|a(k)\Psi\big\|^{2}dk<\infty.
\end{equation*}
\textit{In that  case} 
\begin{equation*}
\big\|\textrm{d}\Gamma_{\textrm{b}}(f)^{1/2}\Psi\big\|^{2}=\displaystyle\int_{\mathbb{R}^{d}}f(k)\big\|a(k)\Psi\big\|^{2}dk.
\end{equation*}
\textbf{\Large{APPENDIX B}}
\\
\hspace{3mm}In this section, we introduce facts about essential self-adjointness and essential spectrum which are used in Section 3 and Section 4.
\vspace{3mm}
\\
For $n\in\mathbb{N}\cup\{0\}$, let $\mathcal{X}_{n}$ be a Hilbert space and  $\mathcal{X}:=\oplus_{n=0}^{\infty}\mathcal{X}_{n}$. Let $\mathcal{X}_{\textrm{fin}}$ be defined  by
\begin{equation*}
\mathcal{X}_{\textrm{fin}}:=\{\Psi=\{\Psi^{(n)}\}_{n=0}^{\infty}\in\mathcal{X}:\exists N\textrm{ such that } \Psi^{(n)}=0 \hspace{3mm}\textrm{for all $n\ge N+1$} \}.
\end{equation*}
The number operator $N_{\mathcal{X}}$ is defined by
\begin{equation*}
D(N_{\mathcal{X}}):=\Big\{\Psi\in \mathcal{X}:\displaystyle\sum_{n=0}^{\infty}n^{2}\big\|\Psi^{(n)}\big\|^{2}_{\mathcal{X}_{n}}<\infty\Big\},
\end{equation*}
\begin{equation*}
(N_{\mathcal{X}}\Psi)^{(n)}:=n\Psi^{(n)},\hspace{5mm}\Psi\in D(N_{\mathcal{X}}),\hspace{5mm}(n\in\mathbb{N}\cup\{0\}).
\end{equation*}
Let $A_{n}$ be a self-adjoint operator on $\mathcal{X}_{n}$, and set $A:=\oplus_{n=0}^{\infty} A_{n}$. Let $B$ be a symmetric operator on $\mathcal{X}$.  We identify $\Psi^{(n)}\in\mathcal{X}_{n}$ as
\begin{equation*}
\Psi^{(n)}=\{0,\cdots,0,\Psi^{(n)},0,\cdots\}\in\mathcal{X}.
\end{equation*}
\textbf{Proposition B.1} [\textit{2}]
\hspace{3mm}\textit{Suppose that following} (1)-(3) \textit{hold:} 
\vspace{2mm}
\\
\hspace{3mm}(1) $\mathcal{X}_{\textrm{fin}}\subset D(B)$ and \textit{$A+B$ is bounded below on} $D(A)\cap \mathcal{X}_{\textrm{fin}}$.
\vspace{2mm}
\\
\hspace{3mm}(2) \textit{There exists an integer $p\in\mathbb{N}$ such that}
\begin{equation*}
\langle \Psi^{(n)},B\Psi^{(m)}\rangle_{\mathcal{X}}=0,\hspace{5mm}\text{whenever}\hspace{3mm}|n-m|\ge p.
\end{equation*}
\vspace{2mm}
\\
\hspace{3mm} (3)
\textit{There exist a constant $c>0$ and a linear operator $L$ on $\mathcal{X}$ such that} $D(((A+B)\upharpoonright D(A)\cap\mathcal{X}_{\textrm{fin}})^{\ast})\subset D(L)$, 
\\
\hspace{10mm} Ran($L\upharpoonright D(L)\cap\mathcal{X}_{n})\subset \mathcal{X}_{n}$ \textit{and}
\begin{equation*}
|\langle \Phi,B\Psi\rangle|\le c\big\|L\Phi\big\|\big\|(N_{\mathcal{X}}+1)^{2}\Psi\big\|, \hspace{3mm}\Psi\in\mathcal{X}_{\textrm{fin}},\hspace{3mm}\Phi\in D(L).
\end{equation*}
\textit{Then $A+B$ is essentially self-adjoint on} $D(A)\cap\mathcal{X}_{\textrm{fin}}$. 
\vspace{3mm}
\\
\hspace{3mm}Let $\mathscr{K}$ and $\mathcal{X}$ be Hilbert spaces. We consider the Hilbert space
 $\mathscr{K}\otimes\mathscr{F}_{\textrm{b}}(\mathcal{X})$.
Let $A$ be a self-adjoint operator on $\mathscr{K}$ and $S$ be a non negative self-adjoint operator on $\mathcal{X}$. Then
\begin{equation*}
H_{0}:=A\otimes1+1\otimes \textrm{d}\Gamma_{\textrm{b}}(S) 
\end{equation*}
is self-adjoint on $D(A\otimes1)\cap D(1\otimes \textrm{d}\Gamma_{\textrm{b}}(S))$. Let $H_{I}$ be a symmetric operator on $\mathscr{K}\otimes \mathscr{F}_{\textrm{b}}(\mathcal{X})$ and 
\begin{equation}
H:=H_{0}+H_{I}.
\end{equation}
Let us recall a notion of weak commutator.
\vspace{3mm}
\\
\textbf{Definition B.2.} [\textit{5}] \hspace{3mm} \textit{Let $\mathscr{X}$ be a Hilbert space. Let $A$ and $B$ be densely defined linear operators on $\mathscr{X}$. If there exists a dense subspace $\mathscr{Y}$ and a linear operator K such that $\mathscr{Y}\subset D(K)\cap D(A)\cap D(A^{\ast})\cap D(B)\cap D(B^{\ast})$ and}
\begin{equation*}
\langle A^{\ast}\psi,B\phi\rangle-\langle B^{\ast}\psi,A\phi\rangle=\langle \psi, K\phi\rangle,\hspace{5mm}\psi, \phi\in \mathscr{Y},
\end{equation*} 
\textit{then we say that the couple $\langle A,B\rangle$ has the weak commutator on $\mathscr{Y}$ defined by}
\begin{equation*}
[A,B]_{\textrm{w},\mathscr{Y}}:= K\upharpoonright \mathscr{Y}.
\end{equation*}
The next proposition gives a sufficient condition for $\langle A,B\rangle$ to have a weak commutator.
\vspace{3mm}
\\
\textbf{Proposition B.3.}[\textit{5}]\hspace{3mm} \textit{Let $\mathcal{X}$ be a Hilbert space and let $\mathcal{D}$ be a dense subspace of $\mathcal{X}$. Let $A$ and $B$ be densely defined linear operators on $\mathcal{X}$ such that $\mathcal{D}\subset D(A)\cap D(B)\cap D(A^{\ast})\cap D(B^{\ast})$. Assume that there exist a densely defined closed linear operator $C$ on $\mathcal{X}$ and a core $\mathcal{E}_{C}$ of C with the following properties:}
\vspace{2mm}
\\
\hspace{3mm}(1) $\mathcal{E}_{C}\subset \mathcal{D}\subset D(C)$.
\vspace{2mm}
\\
\hspace{3mm}(2) \textit{$A$ and $B$ are $C$-bounded on $\mathcal{E}_{C}$}.
\vspace{2mm}
\\
\hspace{3mm}(3) $\mathcal{E}_{C}\subset D(AB)\cap D(BA)$ and $K:=[A,B]\upharpoonright \mathcal{E}_{C}$ is $C$-bounded on $\mathcal{E}_{C}$.
\vspace{2mm}
\\
\hspace{3mm}(4) \textit{$D(A^{\ast}B^{\ast})\cap D(B^{\ast}A^{\ast})$ is dense in $\mathscr{X}$.}
\vspace{2mm}
\\
\textit{Then $K$ is closable with $D(C)\subset D(\overline{K})$ and $\langle A,B\rangle$ has a weak commutator on $\mathcal{D}$ which is given by}
\begin{equation*}
[A,B]_{\textrm{w},\mathcal{D}}=\overline{K}\upharpoonright \mathcal{D}.
\end{equation*}
\textbf{Proposition B.4.}[\textit{5}]
\hspace{3mm}\textit{Suppose that following} (1) and (2) \textit{hold}.
\vspace{2mm}
\\
\hspace{3mm}(1)\textit{$H$ is self-adjoint and bounded below.}
\vspace{2mm}
\\
\hspace{3mm} (2)\textit{For any $u\in D(S)\cap D(S^{-1/2})$, the couple $\langle H_{I}, I\otimes A(u)^{\ast}\rangle$ has the weak commutator} $[ H_{I}, I\otimes A(u)^{\ast}]_{\textrm{w},D(H)}$
\\
\hspace{7mm}\textit{ on }$D(H)$. \textit{Furthermore, for any $\Psi\in D(H)$, and any sequences $\{u_{n}\}_{n=1}^{\infty}\subset D(S)\cap D(S^{-1/2})$ such that 
\\
\hspace{7mm}$\big\|u_{n}\big\|=1$, $\wlim_{n\rightarrow\infty} u_{n}=0$},
\begin{equation*}
\displaystyle\lim_{n\rightarrow\infty}[H_{I}, I\otimes A(u_{n})]_{\textrm{w},D(H)}\Psi=0.
\end{equation*} 
\textit{ If $\sigma(S)=[0, \infty)$, then}
\begin{equation*}
\sigma(H)=\sigma_{\textrm{ess}}(H)=[E_{0}(H),\infty).
\end{equation*}
\textbf{\Large{Acknowledgments}}
\\
\hspace{5mm}The author would like to thank Professor A. Arai of Hokkaido University for useful comments. The author is also grateful to K. Usui, S. Futakuchi, and D. Funakawa for helpful discussions.

\end{document}